\documentclass[twocolumn,appendixfloats,tighten]{aastex6}
\usepackage{newtxtext,newtxmath}
\usepackage[T1]{fontenc}
\usepackage{ae,aecompl}

\usepackage{amsmath}	
\usepackage{amssymb}	

\usepackage{ctable}
\usepackage{url}
\usepackage{xspace}
\usepackage[normalem]{ulem}
\usepackage{hyperref}
\usepackage[all]{hypcap}
\usepackage{graphicx}

\newcommand{\be}{\begin{equation}}
\newcommand{\ee}{\end{equation}}

\def\h2{${\rm\,H_2}$}

\usepackage{color}

\begin{document}

\title{Measuring the delay time distribution of binary neutron stars. I. Through Scaling Relations of the Host Galaxies of Gravitational Wave Events}
\author{Mohammadtaher Safarzadeh\altaffilmark{1,2} \& Edo Berger\altaffilmark{1} }

\altaffiltext{1}{Center for Astrophysics|Harvard \& Smithsonian, 60
  Garden Street, Cambridge, MA, 02138, USA,
  \href{mailto:msafarzadeh@cfa.harvard.edu}{msafarzadeh@cfa.harvard.edu}}
\altaffiltext{2}{School of Earth and Space Exploration, Arizona State
  University, AZ, USA}

\begin{abstract}
The delay time distribution of (DTD) of binary neutron stars (BNS) remains poorly constrained, mainly by the small known population of Galactic binaries, 
the properties of short gamma-ray burst host galaxies, and inferences from $r$-process enrichment.  In the new era of BNS merger detections through gravitational waves (GW), 
a new route to the DTD is the demographics of the host galaxies, localized through associated electromagnetic counterparts.  
This approach takes advantage of the correlation between star formation history (SFH) and galaxy mass, such that the convolution of the SFH and DTD impacts the BNS merger rate as a function of galaxy mass.  
Here we quantify this approach for a power law DTD governed by two parameters: the power law index ($\Gamma$) and a minimum delay time ($t_{\rm min}$).  
Under the reasonable assumption that EM counterparts are likely only detectable in the local universe, accessible by the current generation of GW detectors, we study how many host galaxies at $z\sim 0$ are required to constrain the DTD parameters.  We find that the DTD is mainly imprinted in the statistics of massive galaxies (stellar mass of $M_*\gtrsim 10^{10.5}$ M$_\odot$, comparable to the host galaxy of GW170817). 
Taking account of relevant uncertainties we find that $\mathcal{O}(10^3)$ host galaxies are required to constrain the DTD; for a fixed value of $t_{\rm min}$, 
as done in previous analyses of the DTD, $\mathcal{O}(10^2)$ host galaxies will suffice.  
Such a sample might become available within the next two decades, prior to the advent of third-generation GW detectors.

\end{abstract}

\section{Introduction}

The delay time distribution of binary neutron stars is currently only
weakly constrained, mainly by the statistics of the small known sample
of Galactic binary neutron stars (e.g.,
\citealt{VignaGomez:2018th}), from arguments related
to $r$-process enrichment
\citep[e.g., ][]{Matteucci:2014jta,Safarzadeh:2018ub}, and from the properties
of short gamma-ray burst (SGRB) host galaxies
\citep{Zheng:2007hl,lb10,fbc+13,Berger14}.  The DTD is generally
expected to follow a power law distribution based on the following
reasoning.  After the formation of the BNS, the binary's orbit decays
through the emission of gravitational waves on a timescale that
depends on the binary's separation as $t\propto a^4$, where $a$ is the
semi-major axis of the binary at formation \citep{Peters1964}. The distribution of the
merging times, therefore, depends on the distribution of the initial
orbital separation modeled as $dN/da\propto a^{-\beta}$.  The initial
distribution of the O/B star (the progenitors of the NSs) is assumed
to follow a power law $dN/da\propto a^{-1}$. If the binary goes
through a common envelope phase, then the distribution of the
separation becomes steeper and approaches $dN/da\propto
a^{-3}$. Therefore, the expected merger times follow $dN/dt_{\rm
  merge}\propto t^{\Gamma}$, where $\Gamma\equiv-\beta/4-3/4$. For
those two limiting cases, $\Gamma$ ranges from $-1.5$ to $-1$
\citep[e.g., ][]{Belczynski:2018vr}.  The various weak observational
constraints are roughly in agreement with these values.

Separate from the slope of the power law distribution, the minimum
timescale for BNS mergers ($t_{\rm min}$) is another parameter that is
equally important in determining the merger rate across cosmic time.
From population synthesis models $t_{\rm min}$ could be as short as a
few Myr \citep{Dominik:2012cwa}, but various effects could serve to
set a minimum initial separation that will increase the value of
$t_{\rm min}$.  Observationally, the two key DTD parameters are
approximately degenerate with each other in that it is not trivial to
distinguish between a DTD with a steep slope but larger $t_{\rm min}$
and a DTD with shallow slope but shorter $t_{\rm min}$.  Recent
simulations have shown that fast merging channels are needed to
explain the fraction of all the metal-poor stars that are $r$-process
enriched \citep{Matteucci:2014jta,Safarzadeh:2018ub}, as well as for
$r$-process enrichment of ultra-faint dwarf galaxies
\citep{Safarzadeh:2017dq,Safarzadeh:2019dd}. We note that most
previous attempts to determine the DTD have assumed a small value of
$t_{\rm min}$, leaving $\Gamma$ as the only free parameter.  Such an
assumption is also made in the case of the DTD of Type Ia supernovae
(e.g., $t_{\rm min}\sim 40$ Myr based on the minimum lifetime of stars
that produce white dwarfs; \citealt{Maoz:2012dg,Maoz:2014gd}), leading
to better constraints on the power law index.

The shape of the DTD is also imprinted in the demographics of the
galaxies that host BNS mergers in the local universe, manifested
either as the host galaxies of SGRBs \citep{Berger14}, or as the host
galaxies of gravitational wave events that can in turn be pinpointed
through the detection of associated electromagnetic counterparts
(e.g., kilonovae; \citealt{Abbott:2017it,Coulter:2017ei,Soares2017}).
This is because the star formation histories of galaxies are
determined by their masses, and the convolution of the SFH with the
DTD will therefore impact the mass distribution of BNS merger host
galaxies \citep{Zheng:2007hl,lb10,fbc+13,Artale2019}.  The detection of the BNS
merger GW170817, and the identification of its host galaxy, pave the
way for utilizing this approach to constrain the DTD.  

Here we use galaxy scaling relations to explore the impact of the DTD
on the distribution of BNS merger host galaxies, and explore the number
of events required to constrain the DTD.  The upcoming observing
campaigns with Advanced LIGO/Virgo and the upcoming detectors KAGRA
and IndIGO are expected to yield BNS merger samples of
$\mathcal{O}(10)$, $\mathcal{O}(100)$, and $\mathcal{O}(1000)$ within
the next year, $\sim 5$ years, and $\sim 20$ years, respectively,
before the advent of third-generation GW detectors.  The structure of
this {\it Letter} is as follows: In \S2 we demonstrate how the shape
of the DTD affects the demographics of BNS merger host galaxies in the
local universe; in \S3 we explore the sample size required to
constrain the shape of the DTD; and in \S4 we discuss the caveats
involved in this analysis.  We adopt the Planck 2015 cosmological
parameters \citep{Collaboration:2016bk}: $\Omega_M=0.308$,
$\Omega_\Lambda=0.692$, $\Omega_b=0.048$, and $H_0=0.678$ km s$^{-1}$
Mpc$^{-1}$.

\section{Method}

We can write the BNS merger rate for a galaxy with halo mass, $M_h$, at
$z=0$ as:
\begin{align}
  \dot{n}(M_h)=&\int_{z_b=10}^{z_b=0}
  \lambda\frac{dP_m}{dt}(t-t_b-t_{\rm
    min})\psi(M_h,z_b)\frac{dt}{dz}(z_b)dz_b,
\end{align}
where 
\be \frac{dt}{dz} = \frac{-1}{(1+z) E(z) H_0}, \ee and
\begin{equation}
\label{eq:hubble-flow}
E(z)=\sqrt{{\Omega}_{m,0}(1+z)^3+{\Omega}_{k,0}(1+z)^2+{\Omega}_{\Lambda}(z)}.
\end{equation}
Here, $\psi(M_h,z)$ is the mean star formation history (SFH) of a
galaxy with $M_h$ at $z=0$ parametrized following
\citet{2013MNRAS.428.3121M}.  We integrate the SFH from $z_b=10$ to
$0$ (where the choice of maximum redshift has little impact on the
calculation); $t_b$ is the cosmic time corresponding to $z_b$;
$\lambda$ is the BNS mass efficiency, assumed to be a fixed value of
$10^{-5}M_{\odot}^{-1}$ independent of redshift or environment;
$dP_m/dt$ is the merger time distribution, which we parametrize to follow a power law, $\propto t^{\Gamma}$ with a minimum delay time, $t_{\rm min}$ and not evolving with redshift.
Although the DTD for binary black holes is likely highly dependent on metallicity, the DTD for BNS systems has been argued to be at most weakly dependent on  metallicity \citep{Dominik:2012cwa}.
The mass efficiency is assumed to be constant, although this parameter could be fit for in principle \citep{safarzadeh3g}, 
due to the limited depth of adLIGO it acts as a normalization constant that could be ignored when studying the distribution of host galaxy masses in the local universe.
We note that the delay time refers to the time since
birth of the ZAMS stars and not when the BNS is formed.  We also
impose a maximum delay time of 10 Gyr for our fiducial case, but our
results are not sensitive for a longer maximum delay time.

\begin{figure*}
\setlength{\tabcolsep}{1mm}
\includegraphics[width=.50\linewidth]{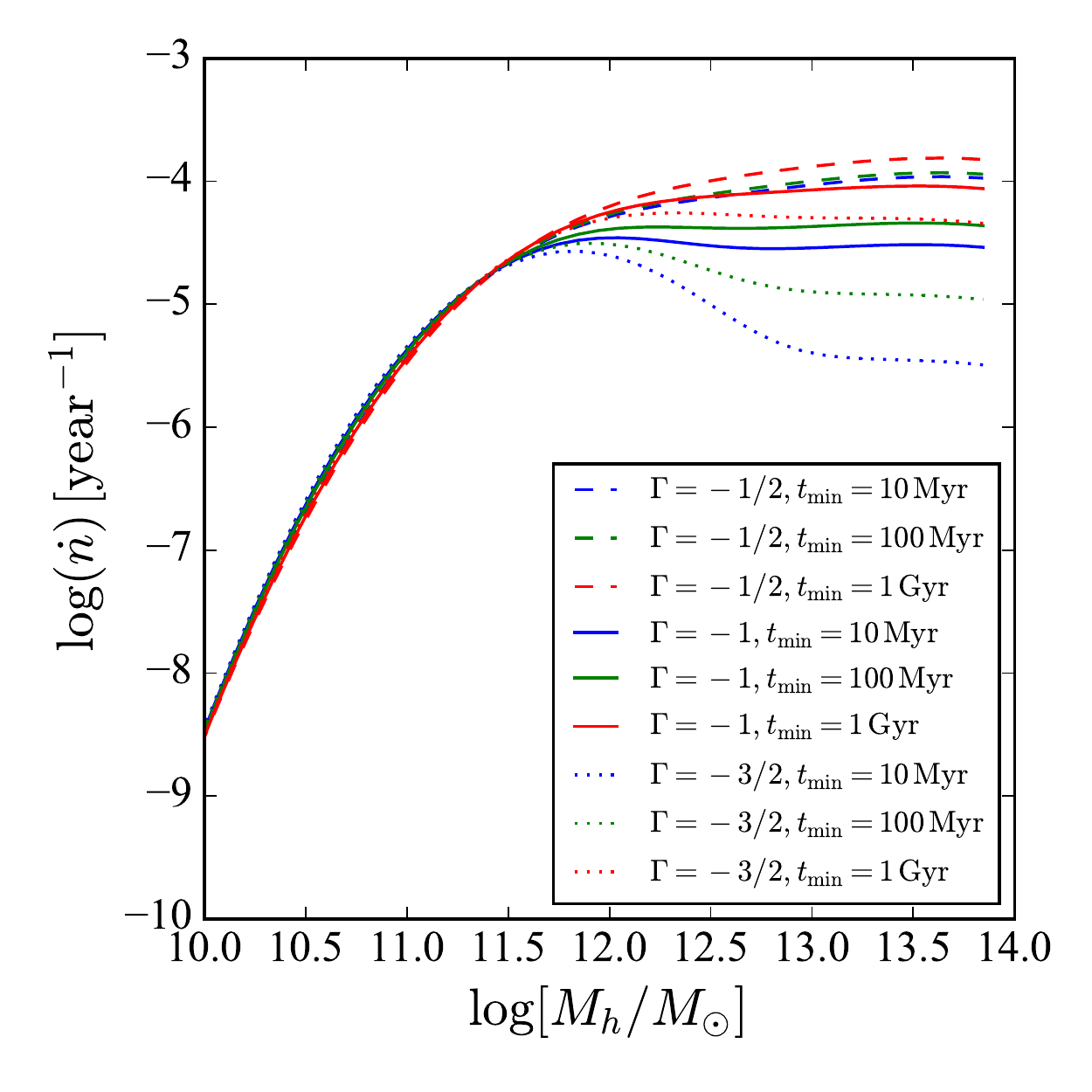}
\includegraphics[width=.50\linewidth]{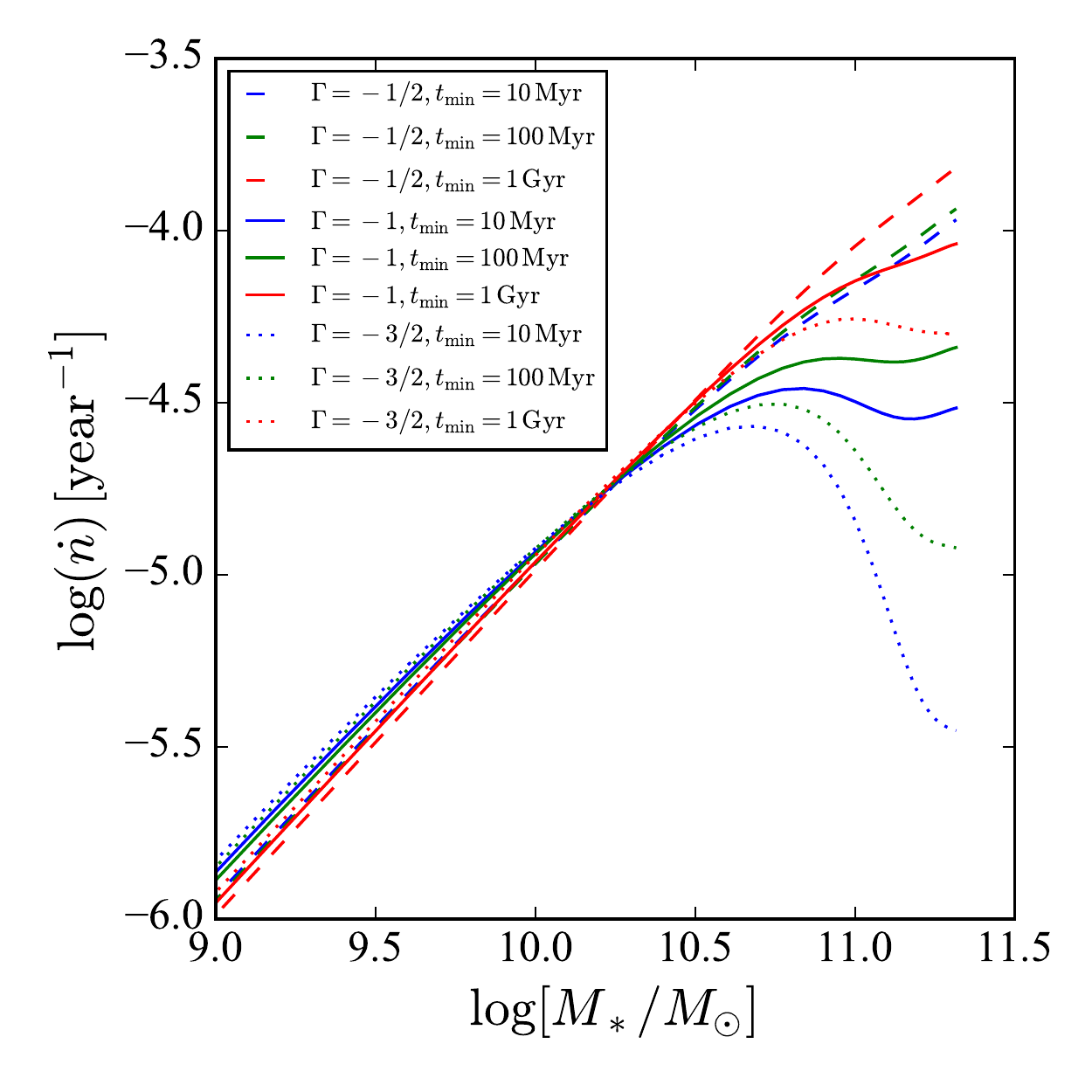}
\includegraphics[width=.50\linewidth]{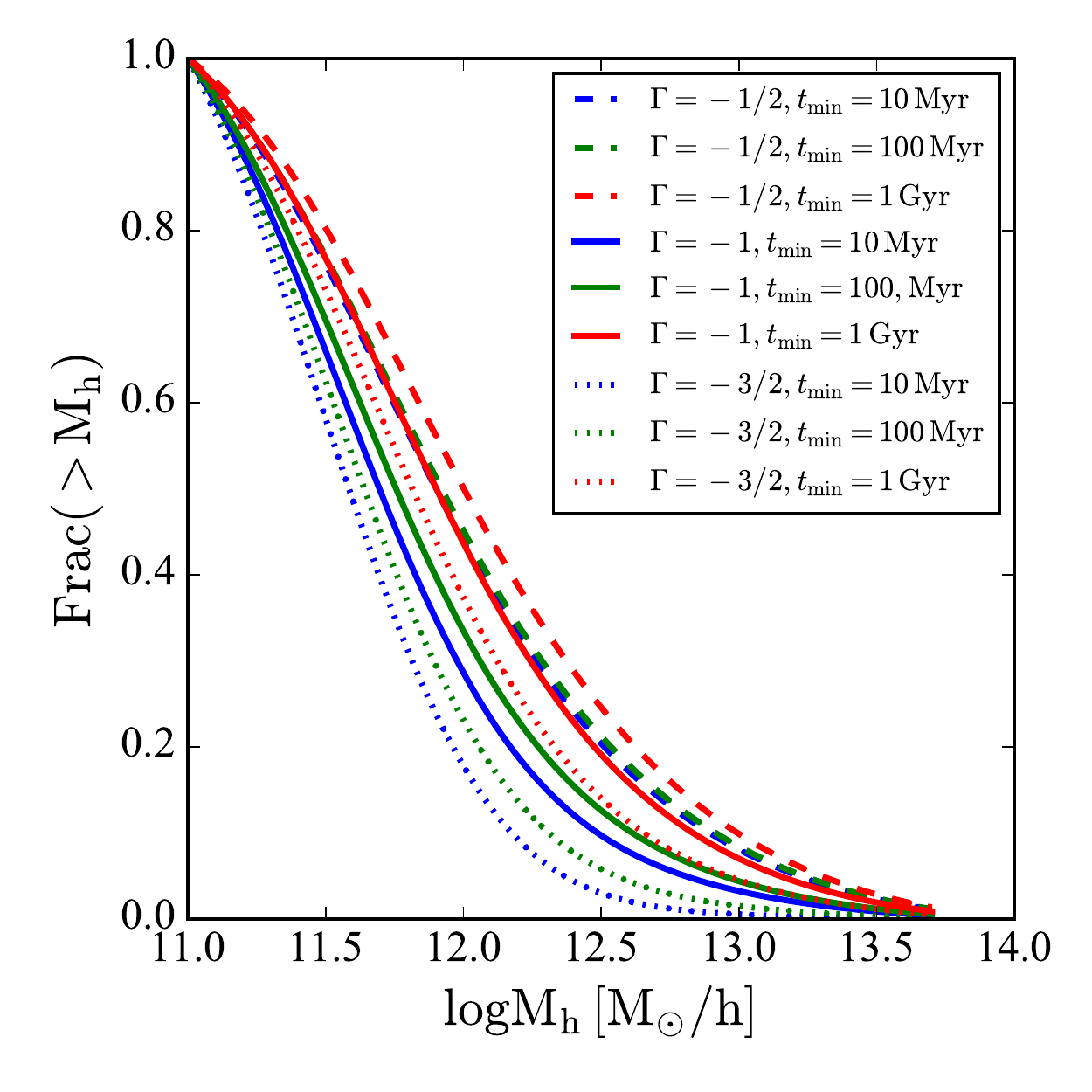}
\includegraphics[width=.50\linewidth]{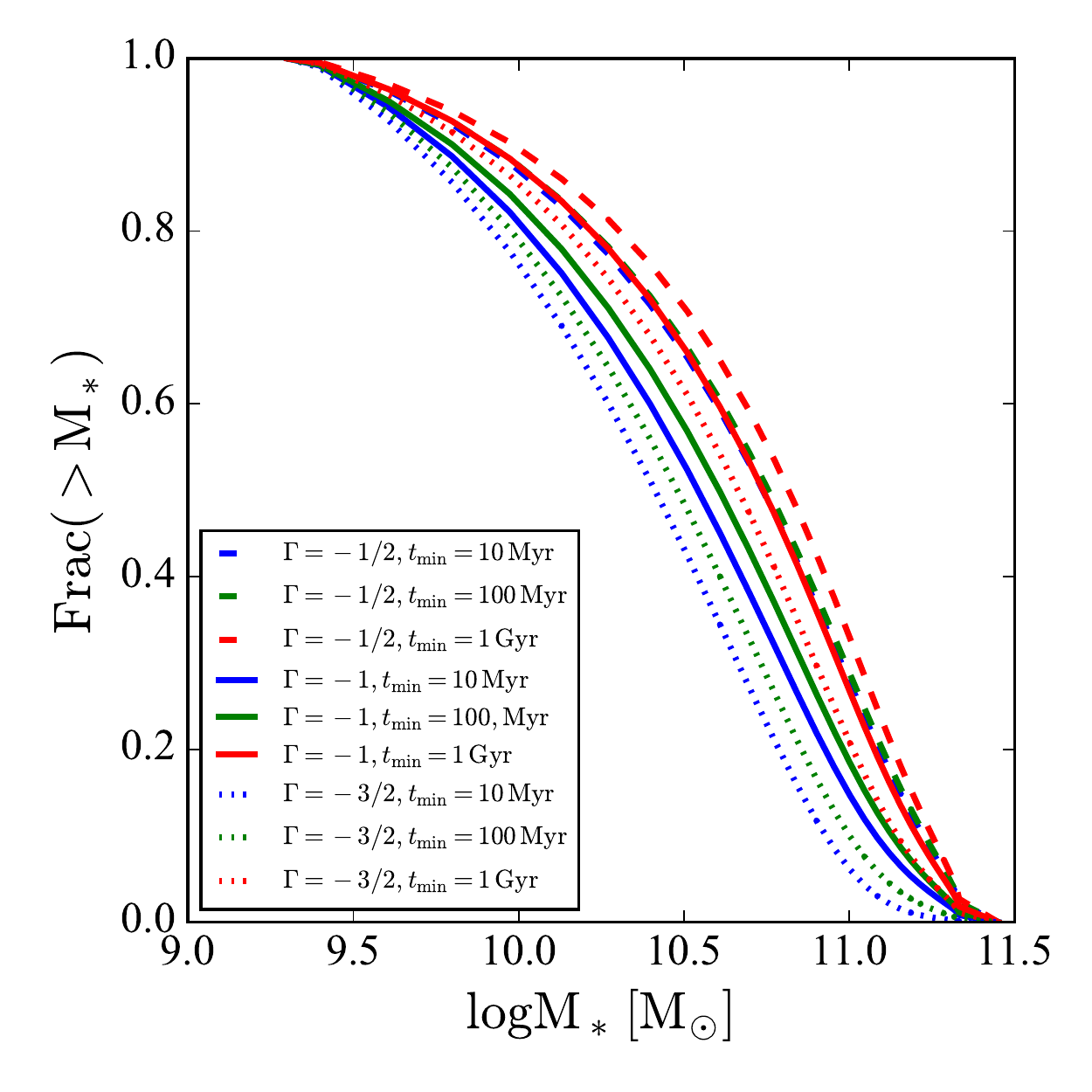}
\caption{{\em Top:} The BNS merger rate ($\dot{n}$) as a function of
  halo mass ({\em Left}) and stellar mass ({\em Right}) at $z=0$ for 9
  different DTDs, with $\Gamma=[-3/2, -1, -1/2]$ and $t_{\rm min}=[10,
  100, 1000]$ Myr.  DTDs with a long delay time (e.g.,
  $\Gamma=-1/2$ and $t_{\rm min}=1000$ Myr) lead to a high merger rate
  in massive galaxies.  {\em Bottom:} The cumulative distribution function of
  BNS merger host galaxies in terms of halo mass  ({\em Left}) and
  stellar mass ({\em Right}) for the same DTDs.  These have been
  constructed by taking into account the halo mass function.}
\label{f:ndot_M_halo}
\end{figure*}

We compute the merger rate per galaxy as a function of halo mass for a
grid of 9 joint choices of $\Gamma=[-1.5, -1.0, -0.5]$ and $t_{\rm
  min}=[10, 100, 1000]$ Myr.  In Figure~\ref{f:ndot_M_halo} we show
the predicted merger rate ($\dot{n}$) as a function of halo mass and
stellar mass for the 9 DTDs.  We find that the key difference between
the various DTDs is apparent at $M_h\gtrsim 10^{12}$ M$_\odot$,
corresponding to $M_*\gtrsim 10^{10.5}$ M$_\odot$.  This is primarily
because on average galaxies of lower masses have fairly flat star
formation histories that are therefore not sensitive to convolution
with the different DTDs.  The high mass galaxies, on the other hand,
have star formation histories that peak at progressively earlier
cosmic time with larger mass.  Therefore, we find that DTDs that favor
long merger timescales (e.g., $\Gamma=-1/2$ and $t_{\rm min}=1$ Gyr)
lead to a higher representation of massive host galaxies.

To determine the observed mass distribution of BNS merger host
galaxies we need to rescale $\dot{n}$ with the halo mass function
(HMF), computed following \citet{Press:1974jb}:
\begin{equation}\label{dndlnM}
  \phi(M_h)\equiv\frac{dn}{dM_h}=\frac{\bar{\rho}}{M_h}f(\nu)\frac{d\nu}{dM_h},
\end{equation} 
where $n$ is the number density of haloes, $\nu$ is the peak-height of
perturbations, $\bar\rho$ is the average density of the universe, and the
first crossing distribution, $f(\nu)$ \citep{Bond:1991cc}, is obtained
from the ellipsoidal collapse model as:
\begin{equation}\label{nufnu}
  \nu f(\nu) = A \sqrt{\frac{a\nu}{2\pi}}\left[1+(a\nu)^{-p}\right]e^{-a\nu/2},
\end{equation}
with $A=0.322$, $p=0.3$, and $a=0.75$ \citep{Sheth:2002fn}.  Here, the
peak height, $\nu$, is defined as $\nu\equiv \delta_{c,0}^2
{\sigma_\chi(R,z)}^{-2},$ with $\delta_{c,0}=1.686$. The variance is
$\sigma^2_\chi(M,z)= \sigma^2_\chi(M,0) D(z)^2,$ with 
\begin{equation}
  \sigma^2_\chi(M,0)= \sigma^2_\chi(R,0)= \int_0^{\infty}\frac{dk}{2
    \pi^2} \,k^2 P_\chi(k) w^2(kR)\,
\label{eqsigM} 
\end{equation}
where $M = 4 \pi R^3 \Omega_M \rho_c/3$, $w(kR)\equiv 3 j_1(kR)/(kR),$
with $j_1(x)\equiv(\sin x-x\cos x)/x^2,$ and $D(z)$ is the linear
growth factor
\begin{equation}
  D(z)\equiv\frac{H(z)}{H(0)} \int_z^{\infty}
  \frac{dz^\prime (1+z^\prime)}{H^3(z^\prime)} \Bigg [\int_0^{\infty}
  \frac{dz^\prime (1+z^\prime)}{H^3(z^\prime)}\Bigg]^{-1}. 
\end{equation}

We compute the cumulative fraction of the BNS merger host halos with
mass above $M_h$ as:
\begin{equation}
f(>M_h) = \frac{\int_{M_h}^{M_{h,\rm max}}\phi(M_h^\prime)
  \dot{n}(M_h^\prime) dM_h^\prime}{\int_{M_{h,\rm min}}^{M_{h,\rm max}}\phi(M_h^\prime) \dot{n}(M_h^\prime)
  dM_h^\prime}, 
\end{equation}
where we consider halos in the range $M_{h,\rm min}=10^{11}$ to
$M_{h,\rm max}=10^{14}$ M$_\odot$.  In Figure~\ref{f:ndot_M_halo} we
show the cumulative distribution function (CDF) for BNS merger host
galaxies as a function of $M_h$ and $M_*$ for the 9 DTDs.  We find
that there is a difference of about 0.5 dex in the median value of
$M_h$ for the range of DTDs, and about 0.7 dex in the value of $M_h$
for the top 20\% most massive galaxies; at the low mass end the CDFs
converge.  Similarly, in terms of stellar mass we find a nearly order
of magnitude spread in the median value of $M_*$.  In both CDFs we
again find a clear degeneracy between $\Gamma$ and $t_{\rm min}$, such
that a DTD with a shallower power law index and small value of $t_{\rm
  min}$ is similar in shape to one with a steeper power law index and
a large $t_{\rm min}$.

Finally, to determine the sample size needed to constrain the DTD we
draw from the constructed CDF of a given pair of [$\Gamma$, $t_{\rm
  min}$], and perform a Kolmogorov-Smirnov (KS) test against other
possible CDFs constructed in a $10\times 10$ interpolated 2D plane of
$\Gamma-t_{\rm min}$.  We use a threshold $P<0.01$ to consider the
CDFs as being drawn from different underlying distributions.  For each
case we repeat the KS test 10 times and determine the median value of
$P$.

We note that the overall approach requires an identified EM counterpart and host galaxy; with only two examples to date (GW170817 and likely S190425z; \citealt{Hosseinzadeh19}) it is difficult to assess the counterpart identification success rate. We show in \citet{safarzadeh3g} how future third-generation GW detectors could enable a measurement of the DTD without the need for EM counterparts. Moreover, we focused on BNS systems formed in the field, while there are other proposed mechanism for their formation, although those are expected to be a minor contributor (e.g., \citealt{Grindlay2006NatPh,Lee2010}).

\section{Results}

\begin{figure*}
\setlength{\tabcolsep}{1mm}
\hspace{0cm}
\includegraphics[width=.33\linewidth]{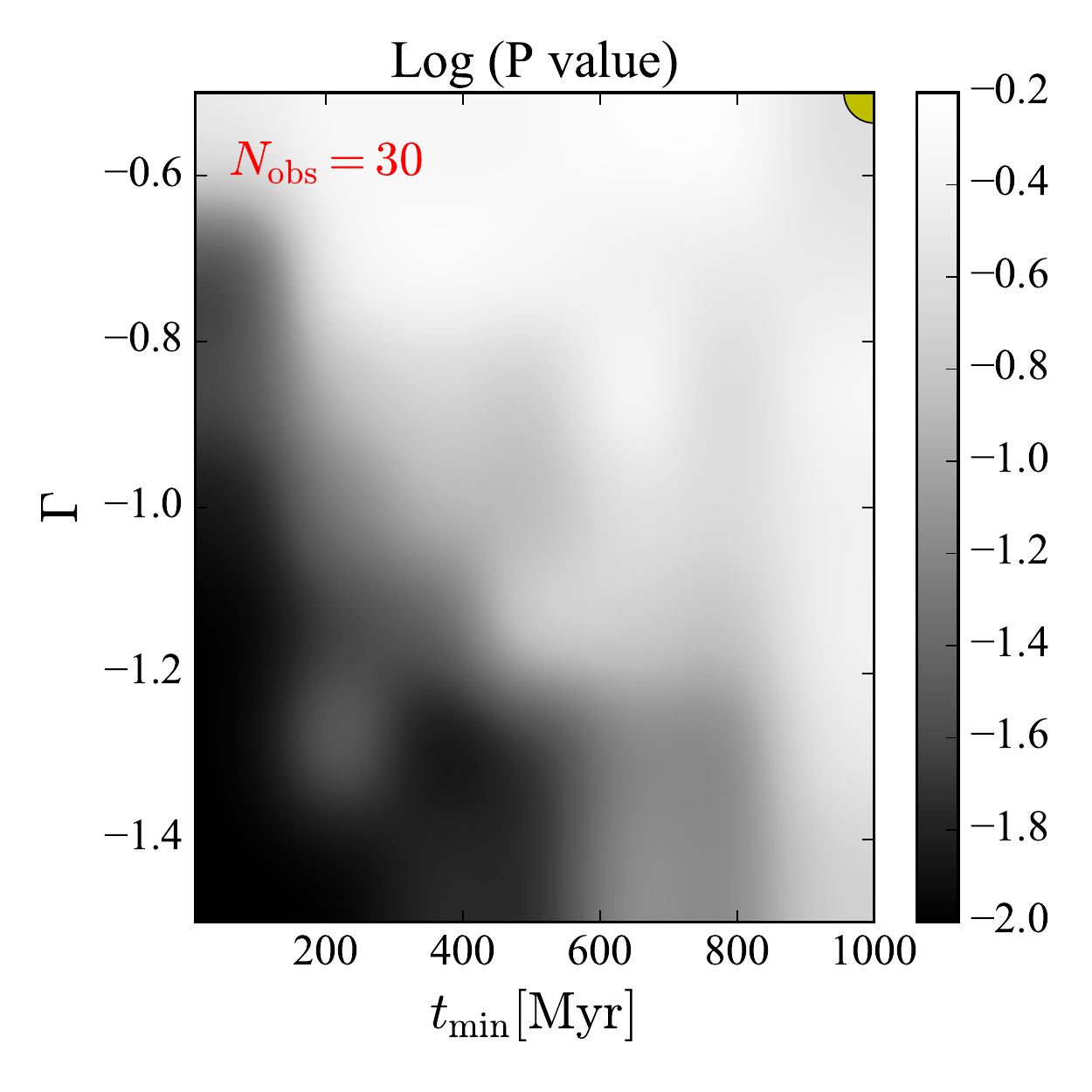}
\includegraphics[width=.33\linewidth]{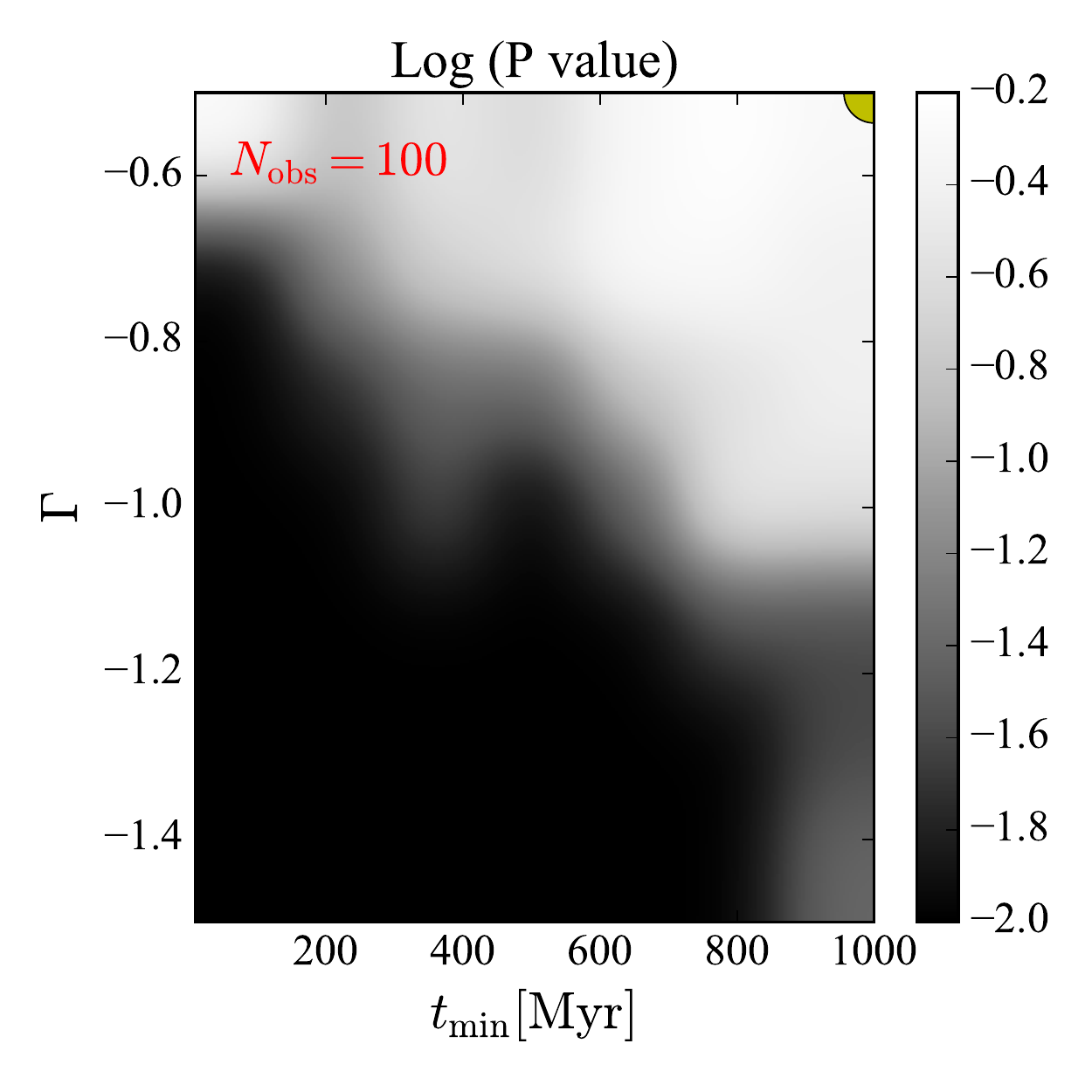}
\includegraphics[width=.33\linewidth]{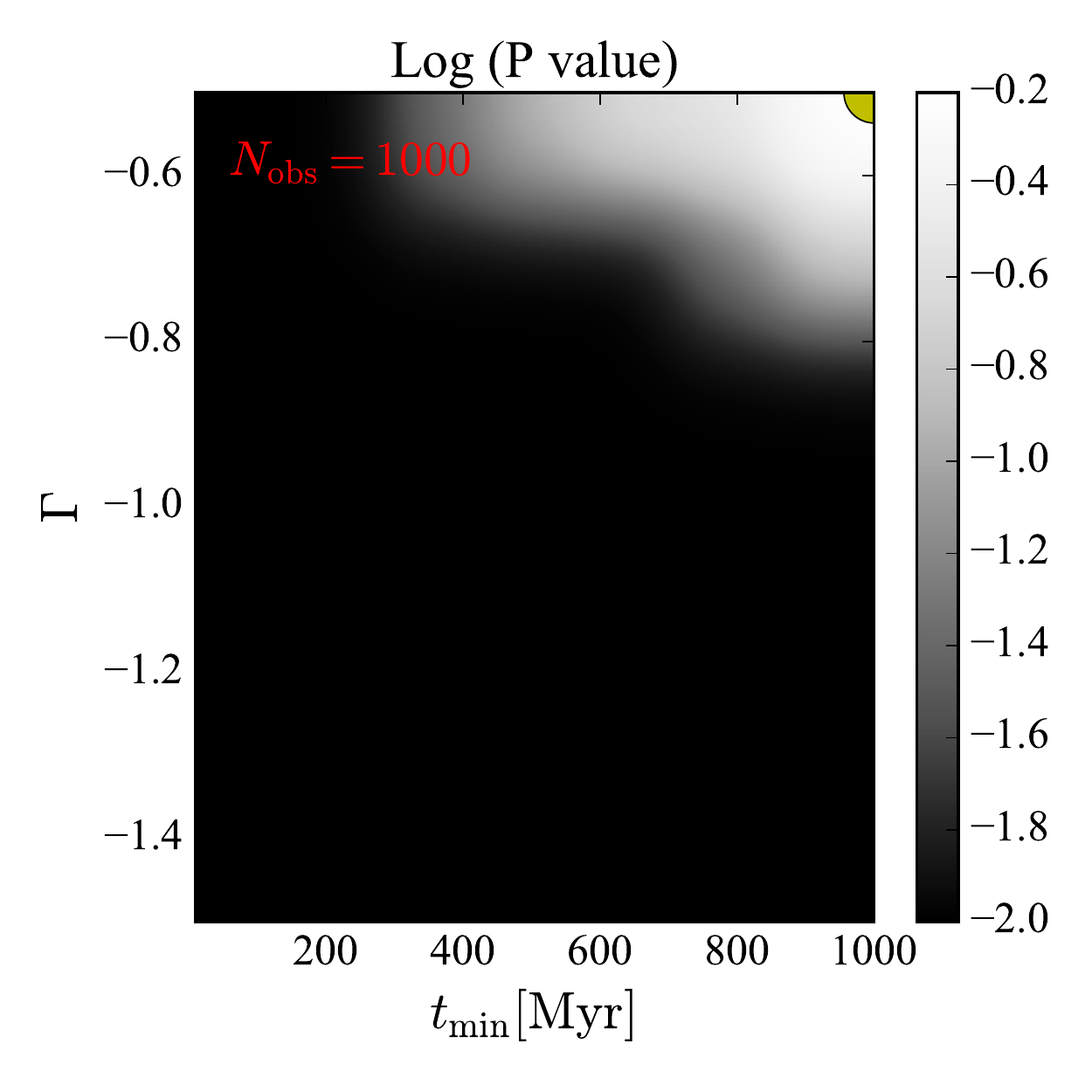}
\includegraphics[width=.33\linewidth]{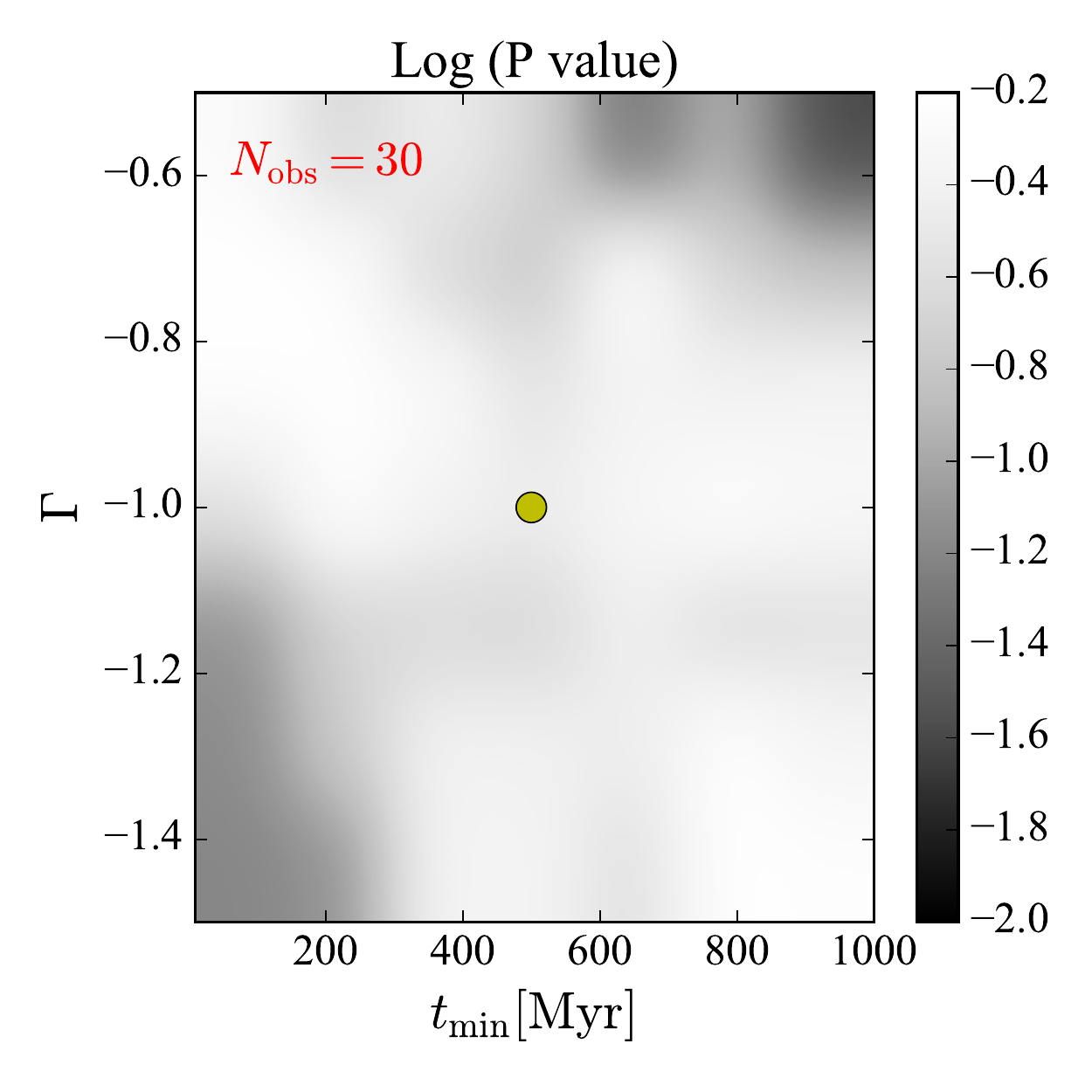}
\includegraphics[width=.33\linewidth]{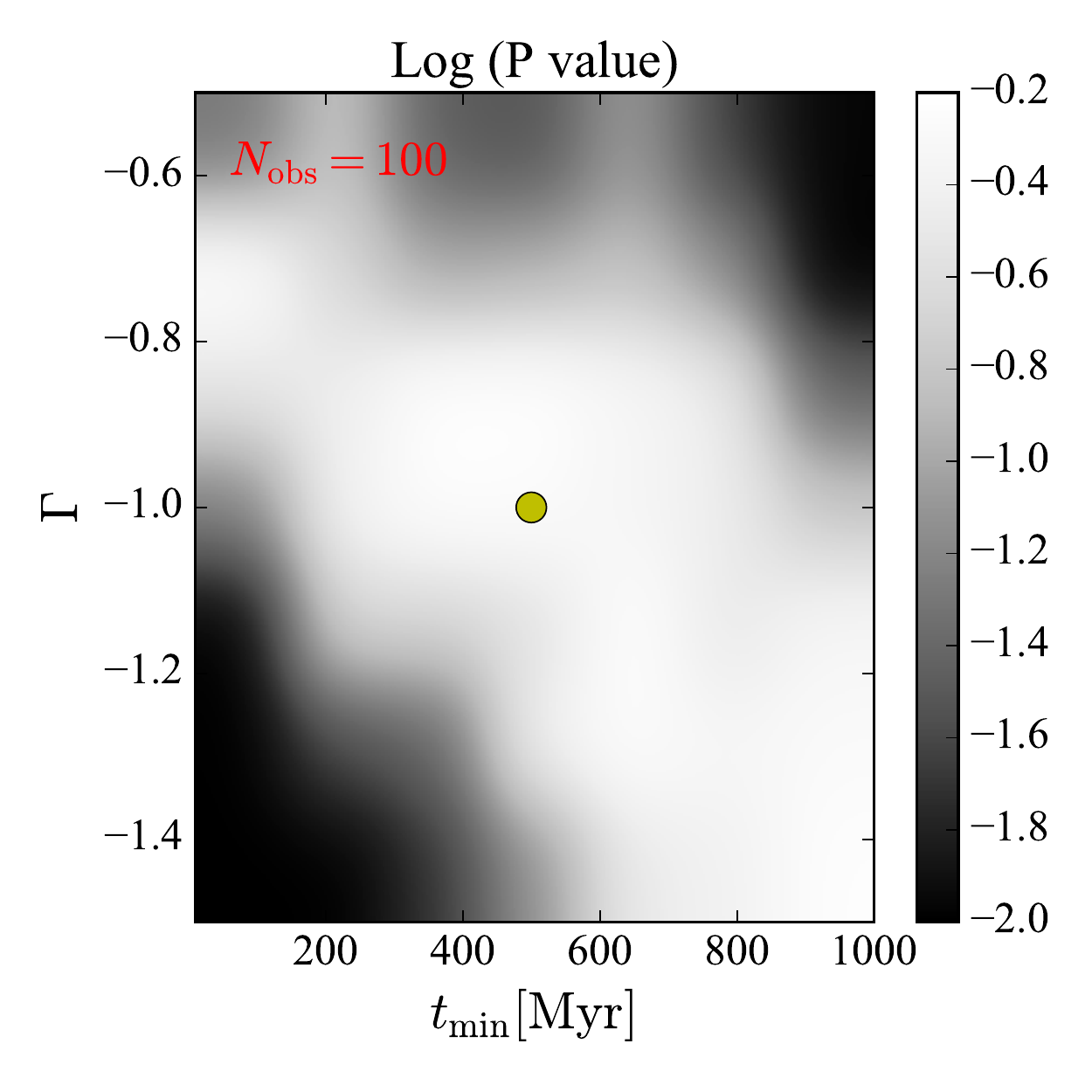}
\includegraphics[width=.33\linewidth]{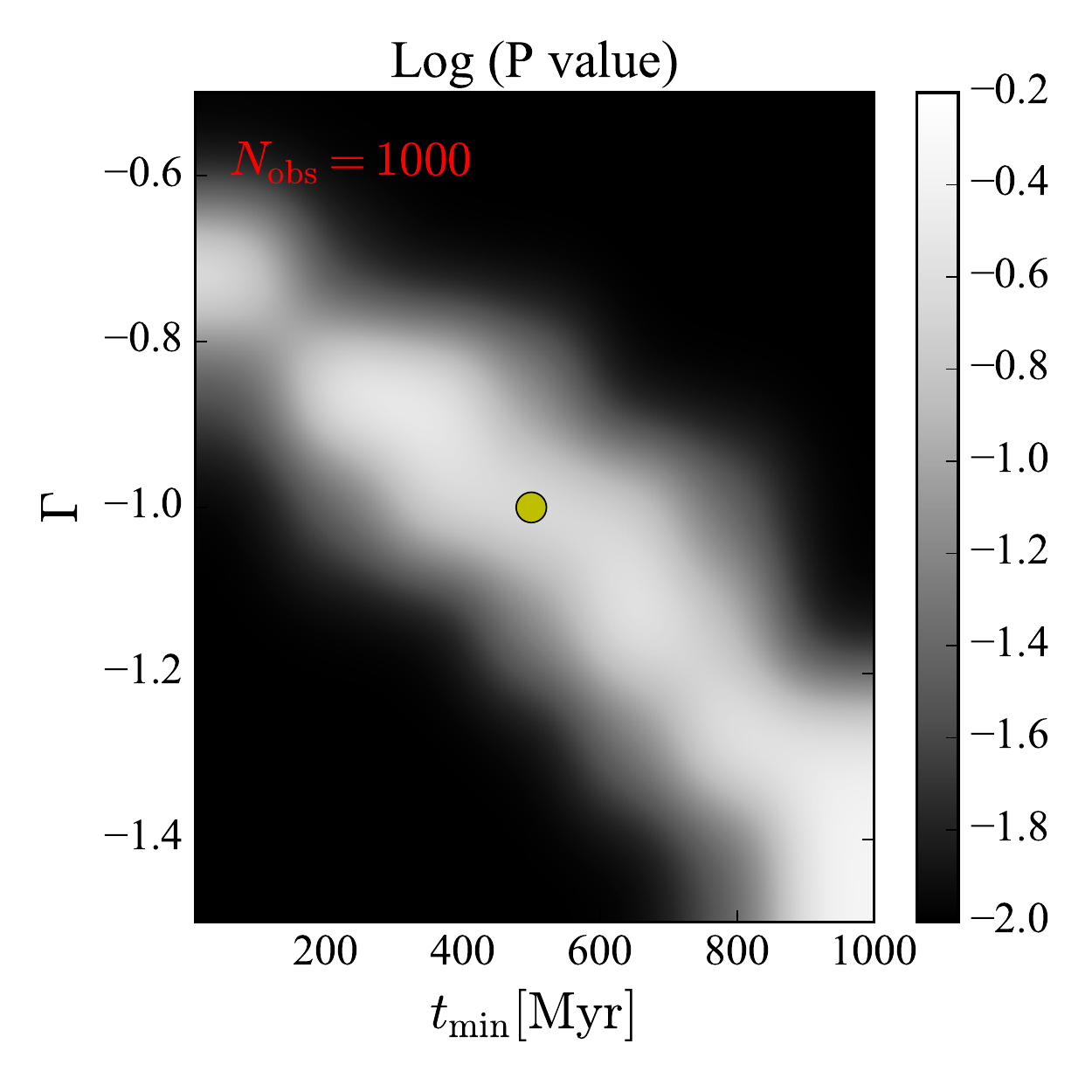}
\includegraphics[width=.33\linewidth]{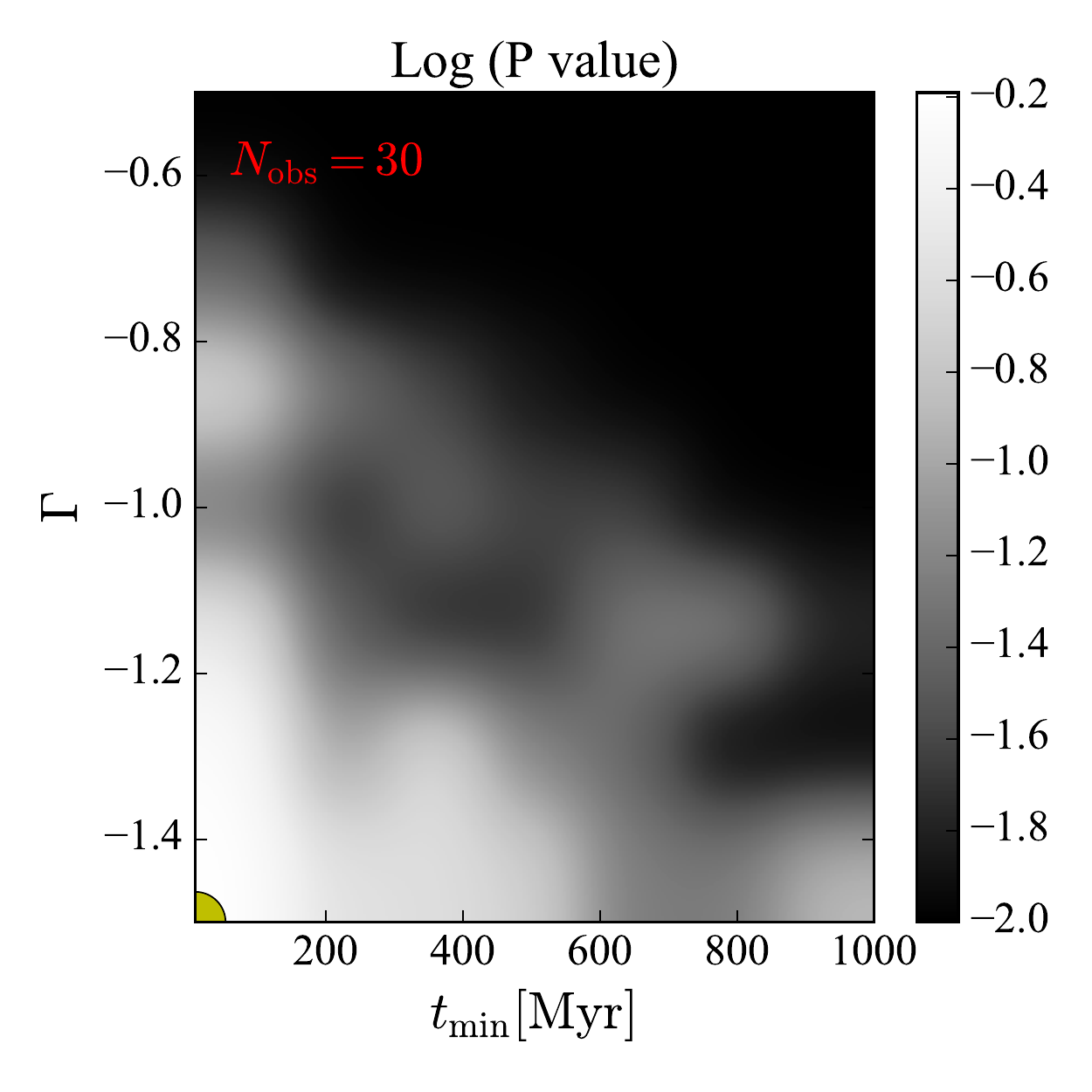}
\includegraphics[width=.33\linewidth]{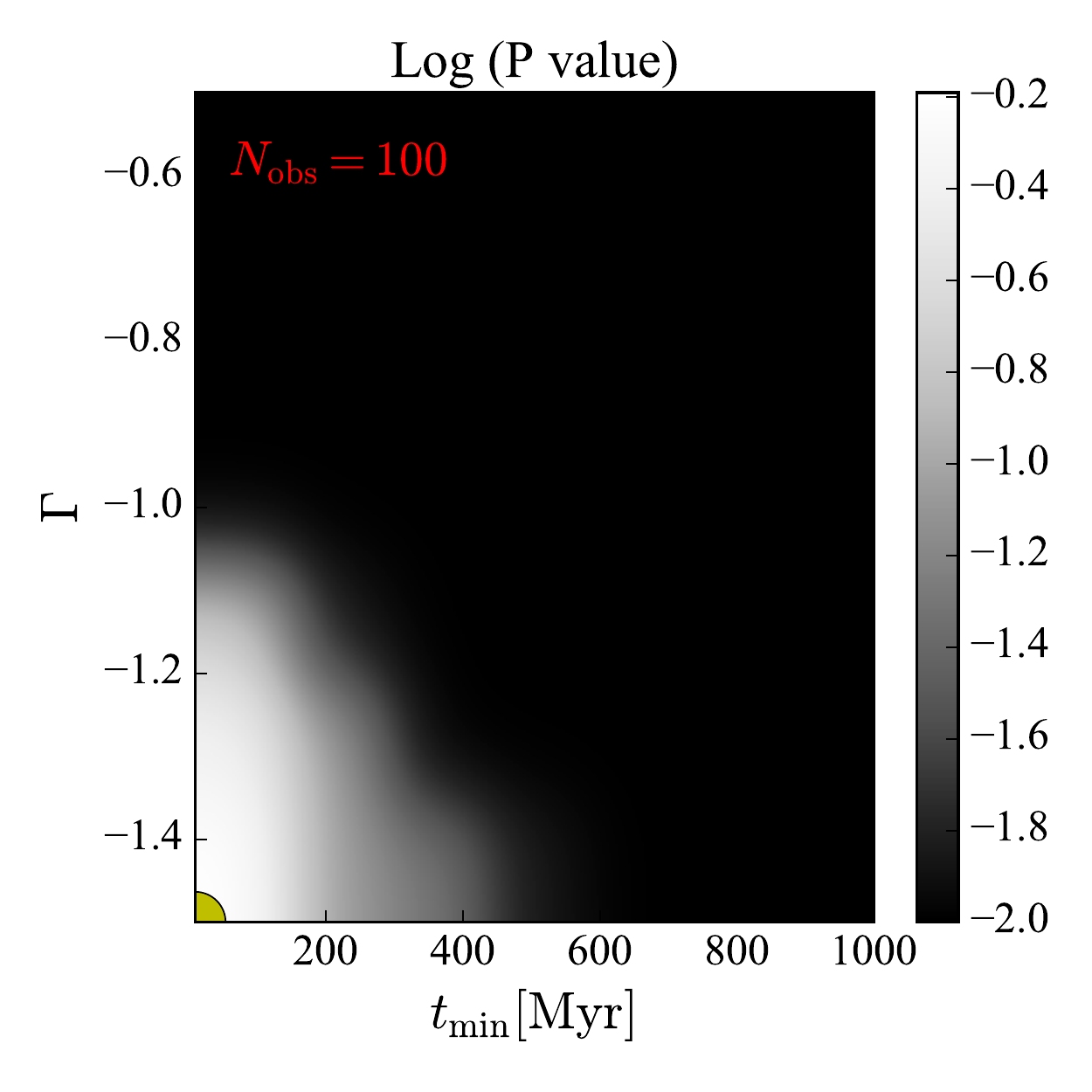}
\includegraphics[width=.33\linewidth]{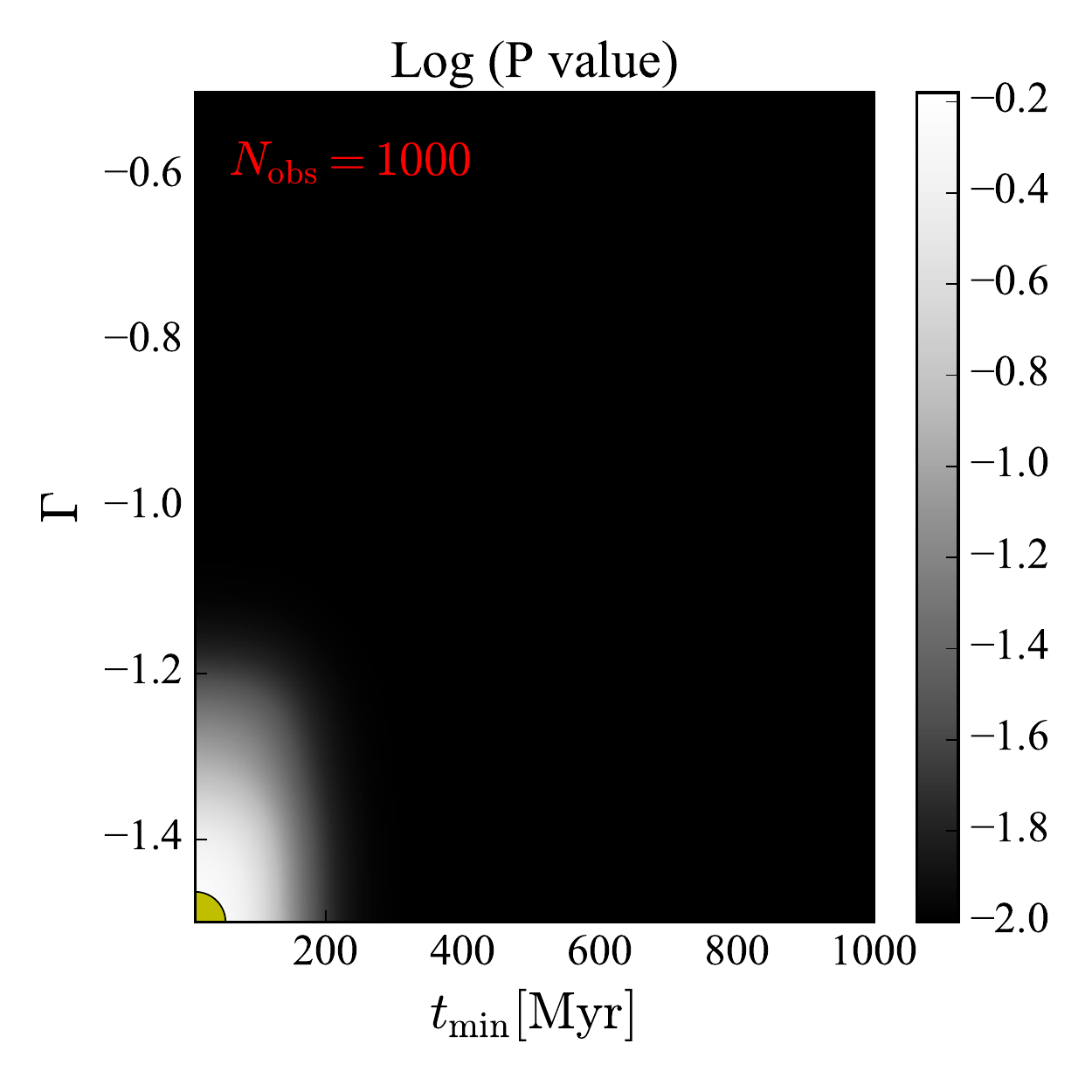}
\caption{Example KS test $P$ value maps as a function of the number of
  observed BNS merger host galaxies, in the space of halo mass. In
  each row the injected model is shown as a yellow circle, and the
  greyscale indicates the $P$ value in the full range of considered
  $\Gamma-t_{\rm min}$ parameter space (darker color indicates lower
  $P$ value).  The columns are for sample of 30, 100, and 100 BNS
  merger host  galaxies.  We find that a sample size of
  $\mathcal{O}(10^3)$ is required to well constrain both $\Gamma$ and
  $t_{\rm min}$, but that some intrinsic degeneracy remains.}
\label{f:final}
\end{figure*}

\begin{figure*}
\setlength{\tabcolsep}{1mm}
\hspace{0cm}
\includegraphics[width=.33\linewidth]{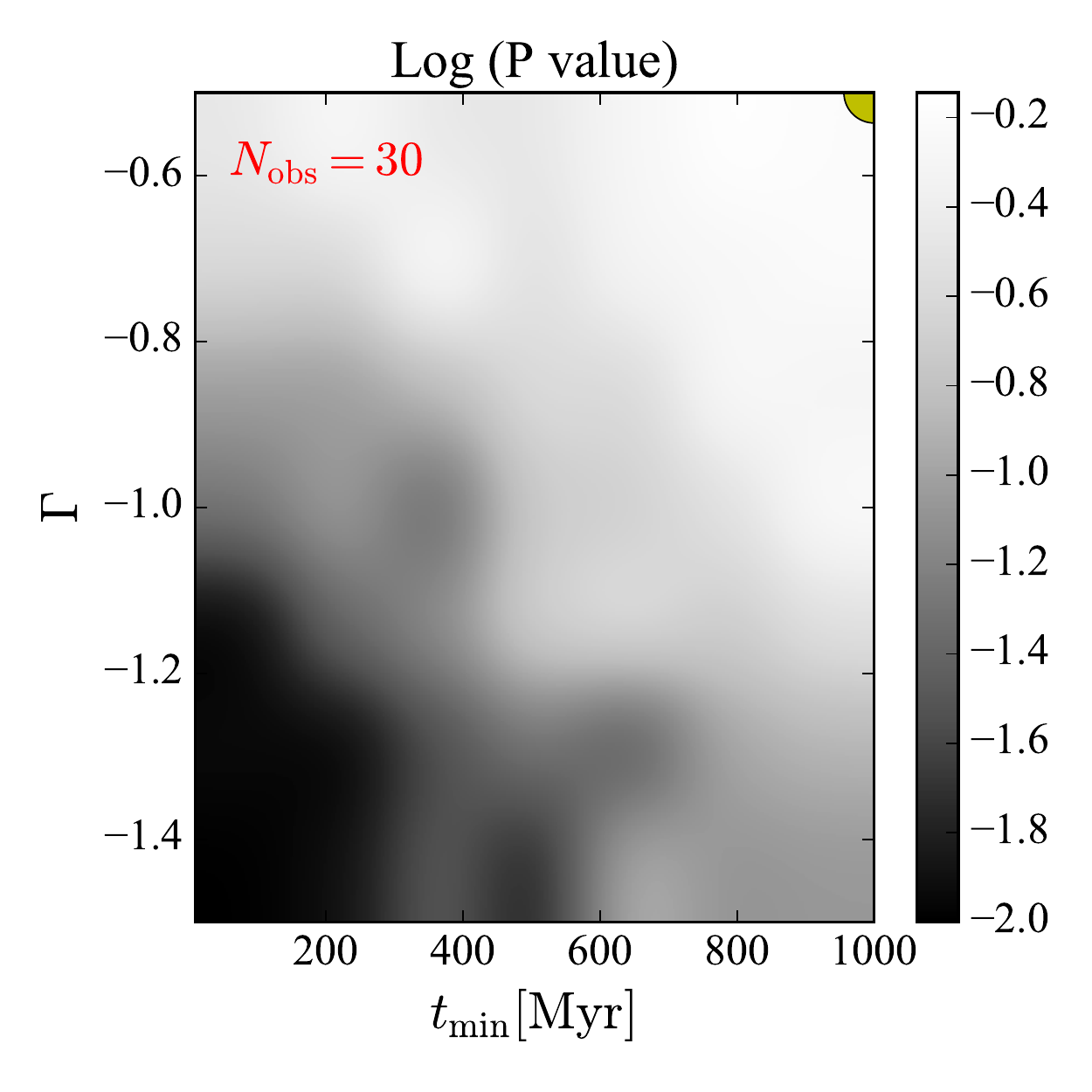}
\includegraphics[width=.33\linewidth]{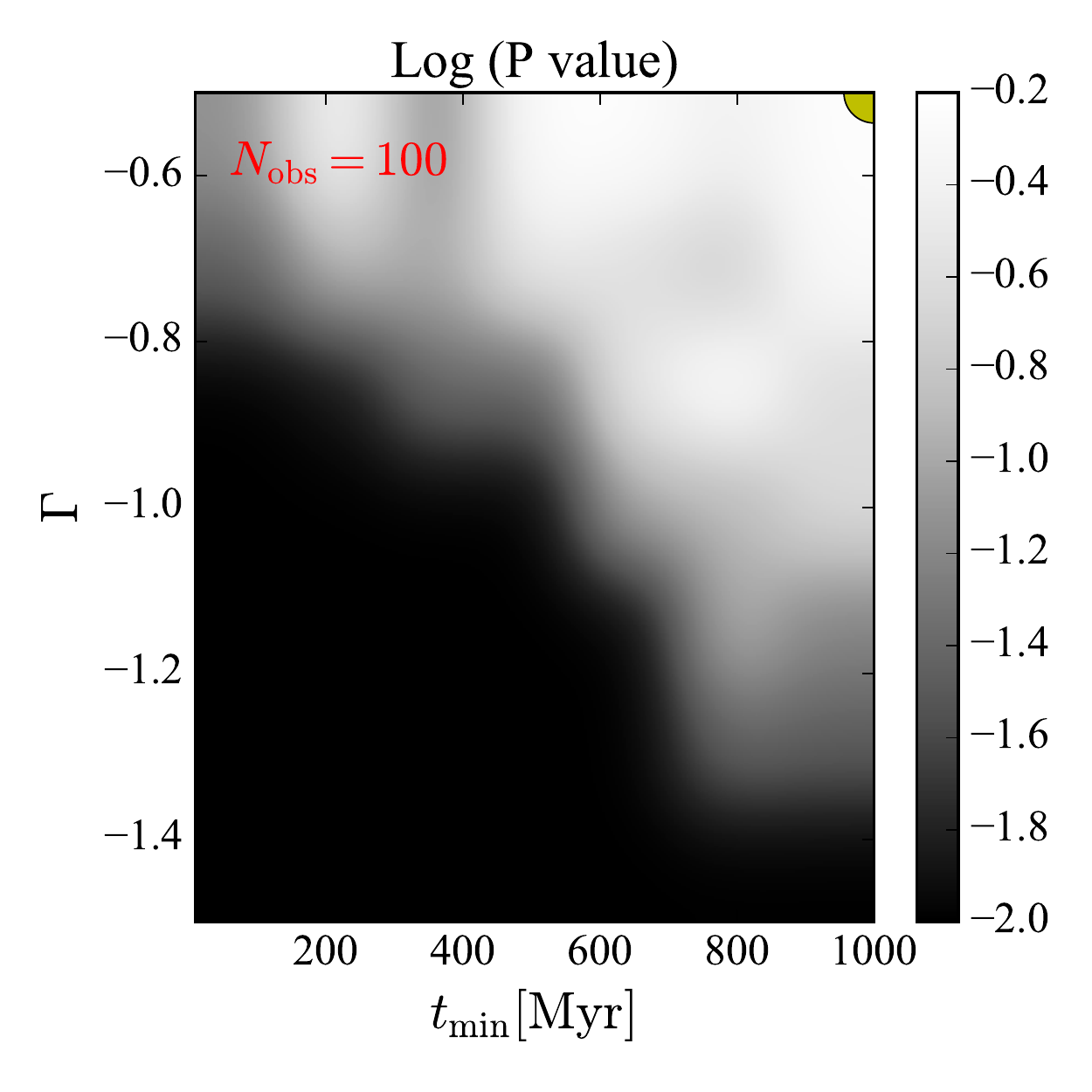}
\includegraphics[width=.33\linewidth]{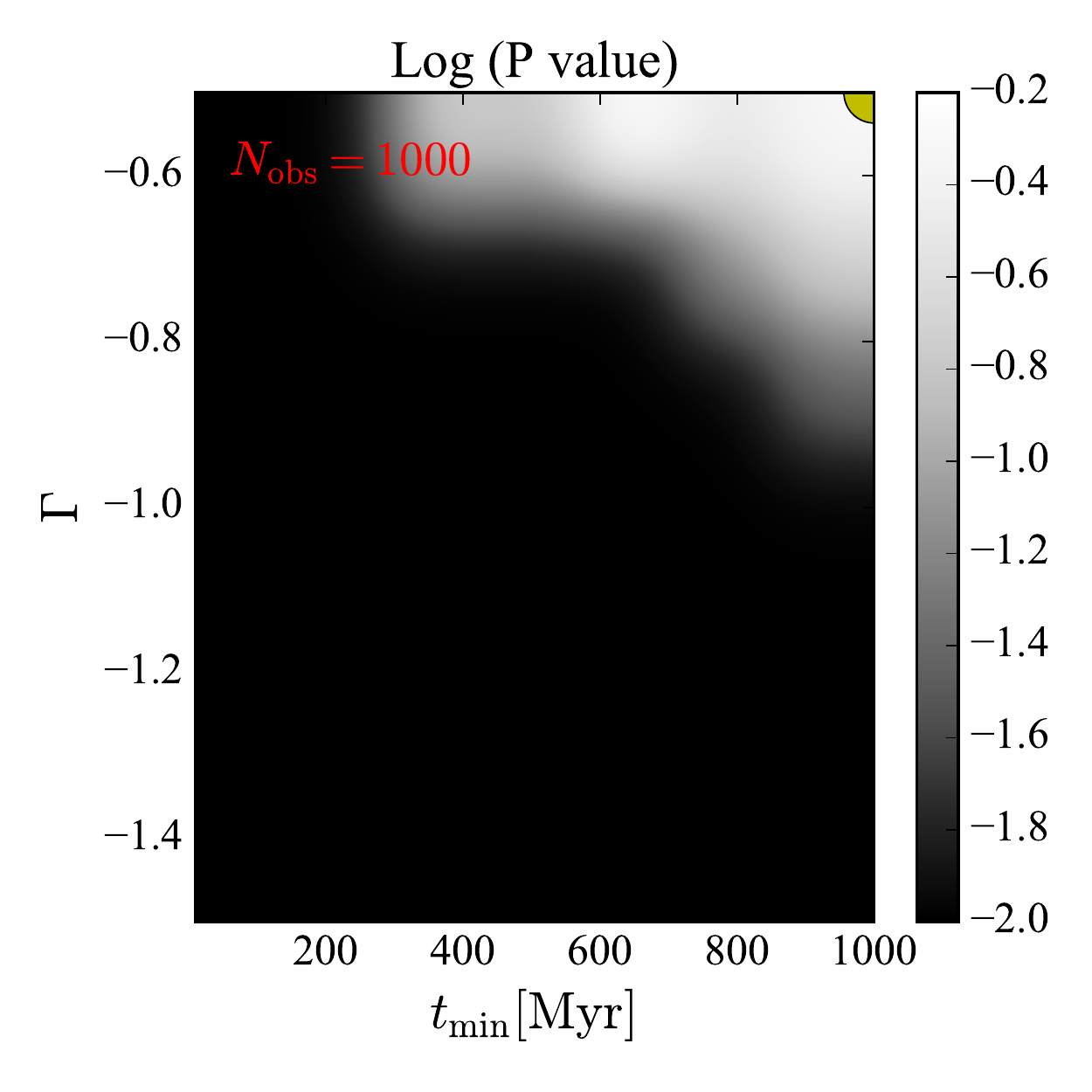}
\includegraphics[width=.33\linewidth]{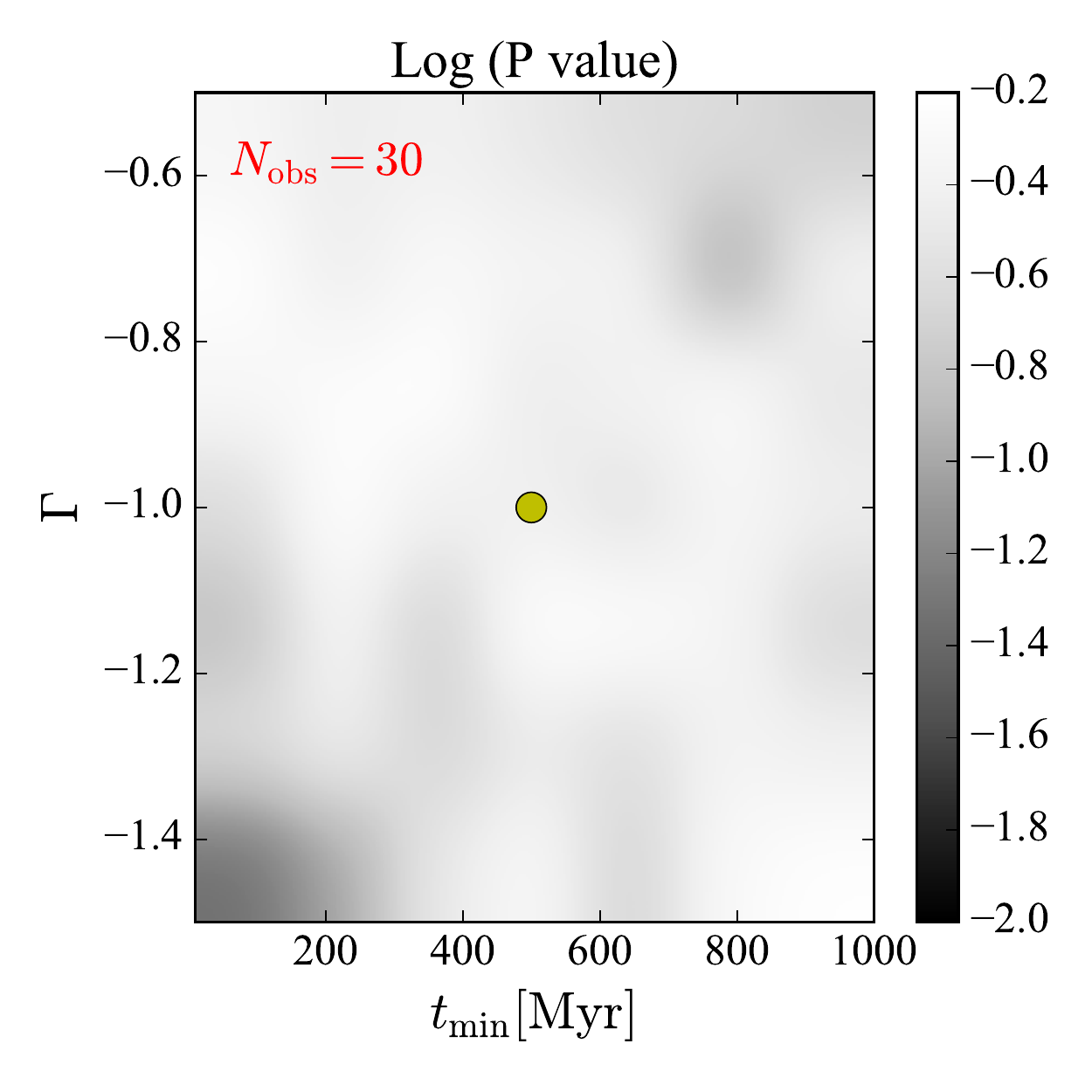}
\includegraphics[width=.33\linewidth]{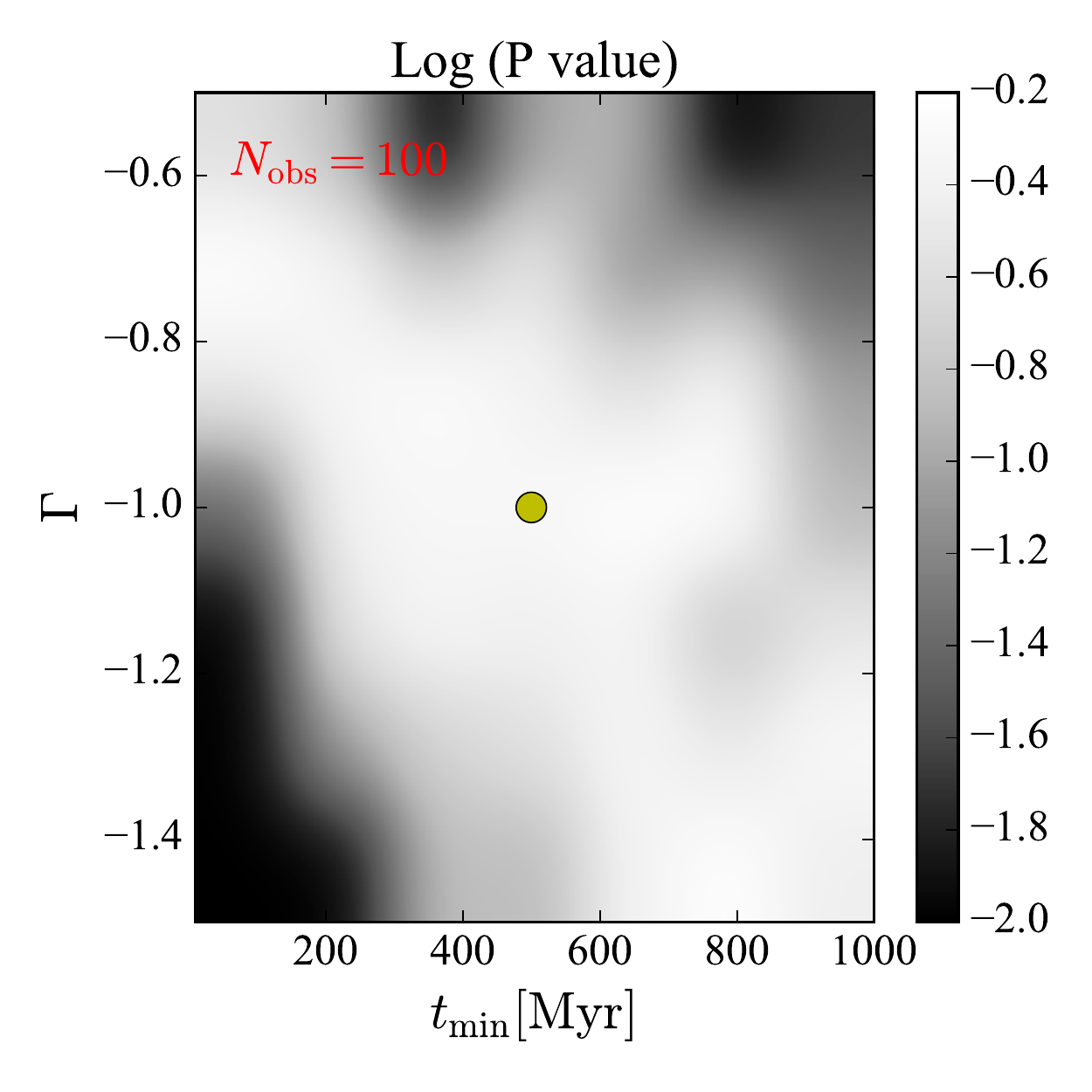}
\includegraphics[width=.33\linewidth]{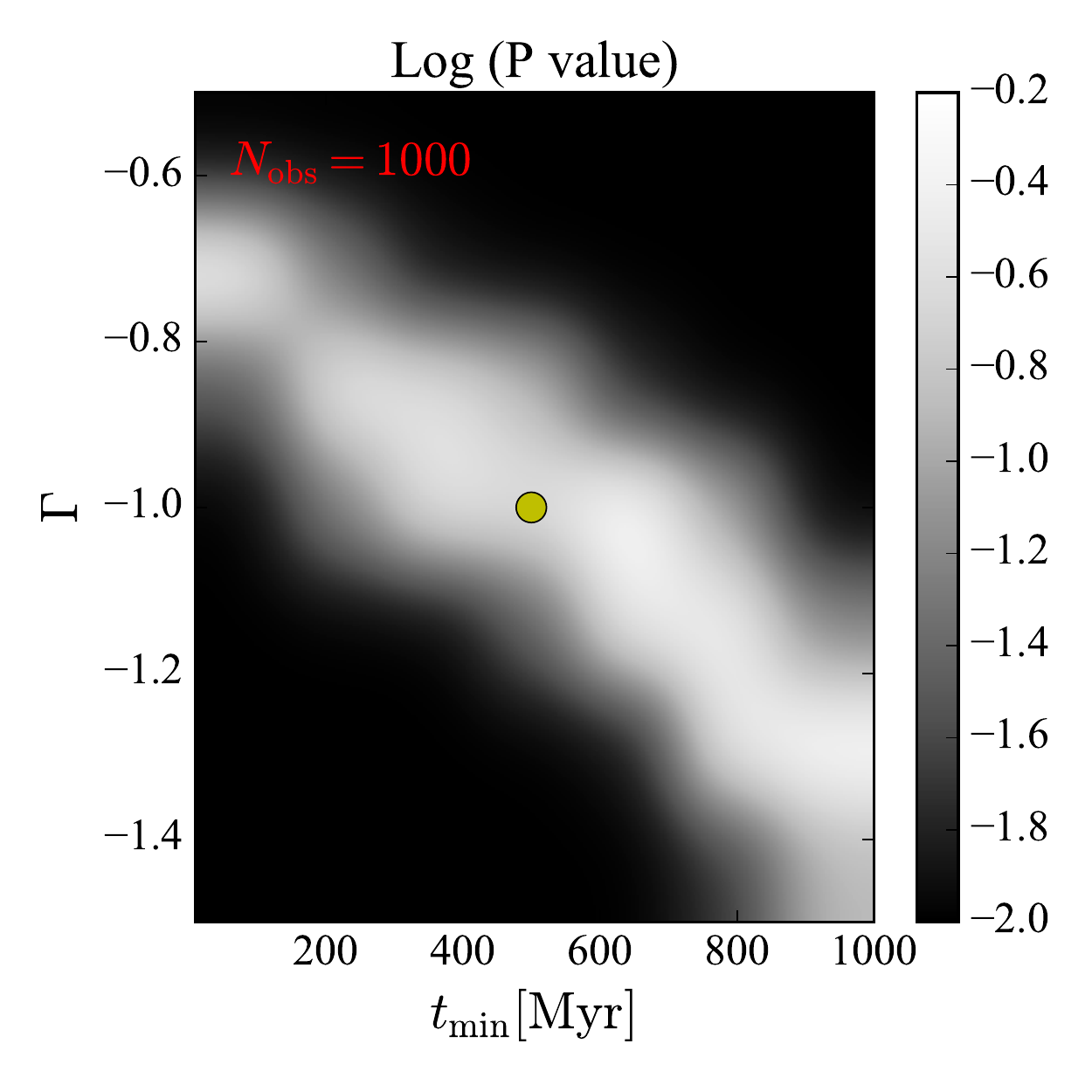}
\includegraphics[width=.33\linewidth]{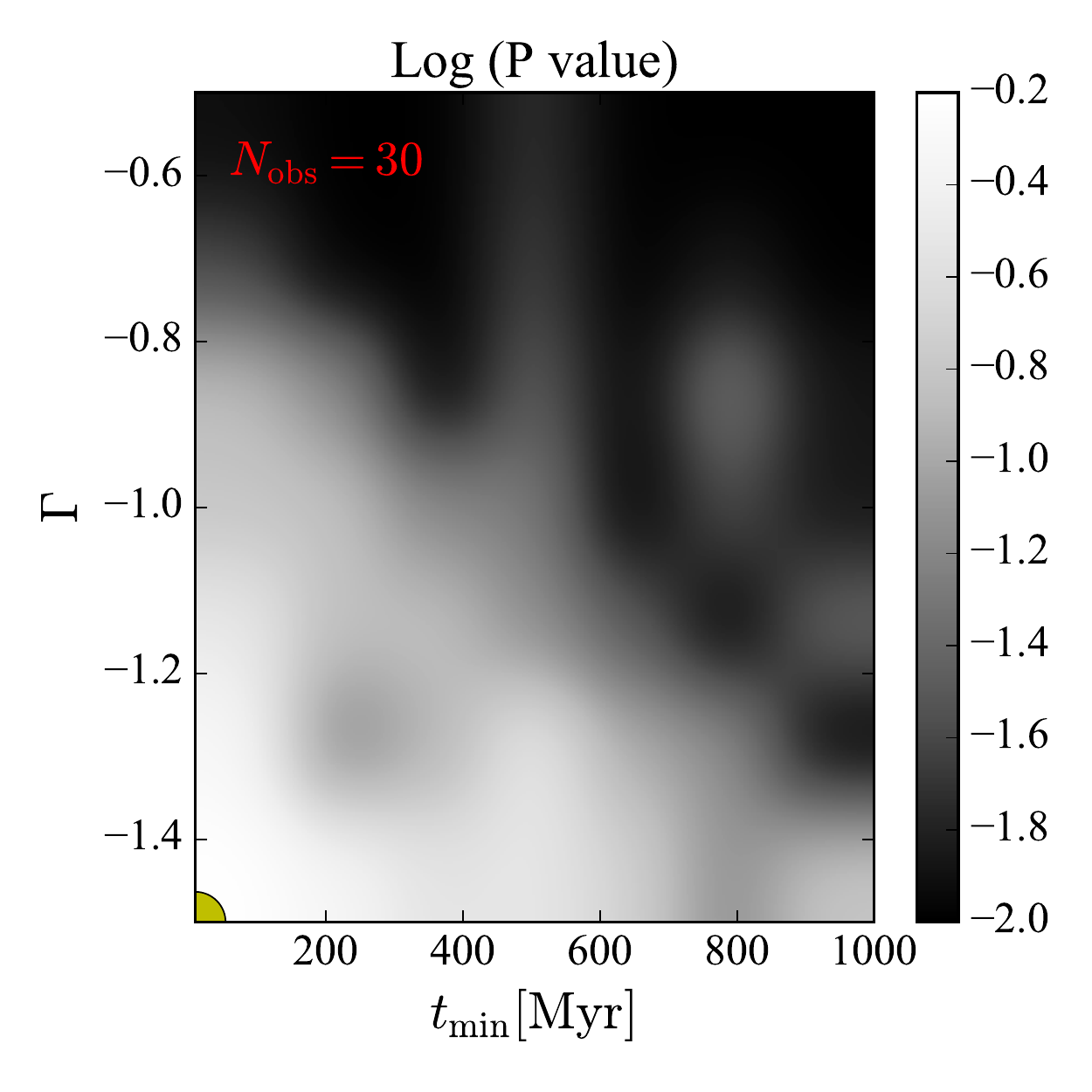}
\includegraphics[width=.33\linewidth]{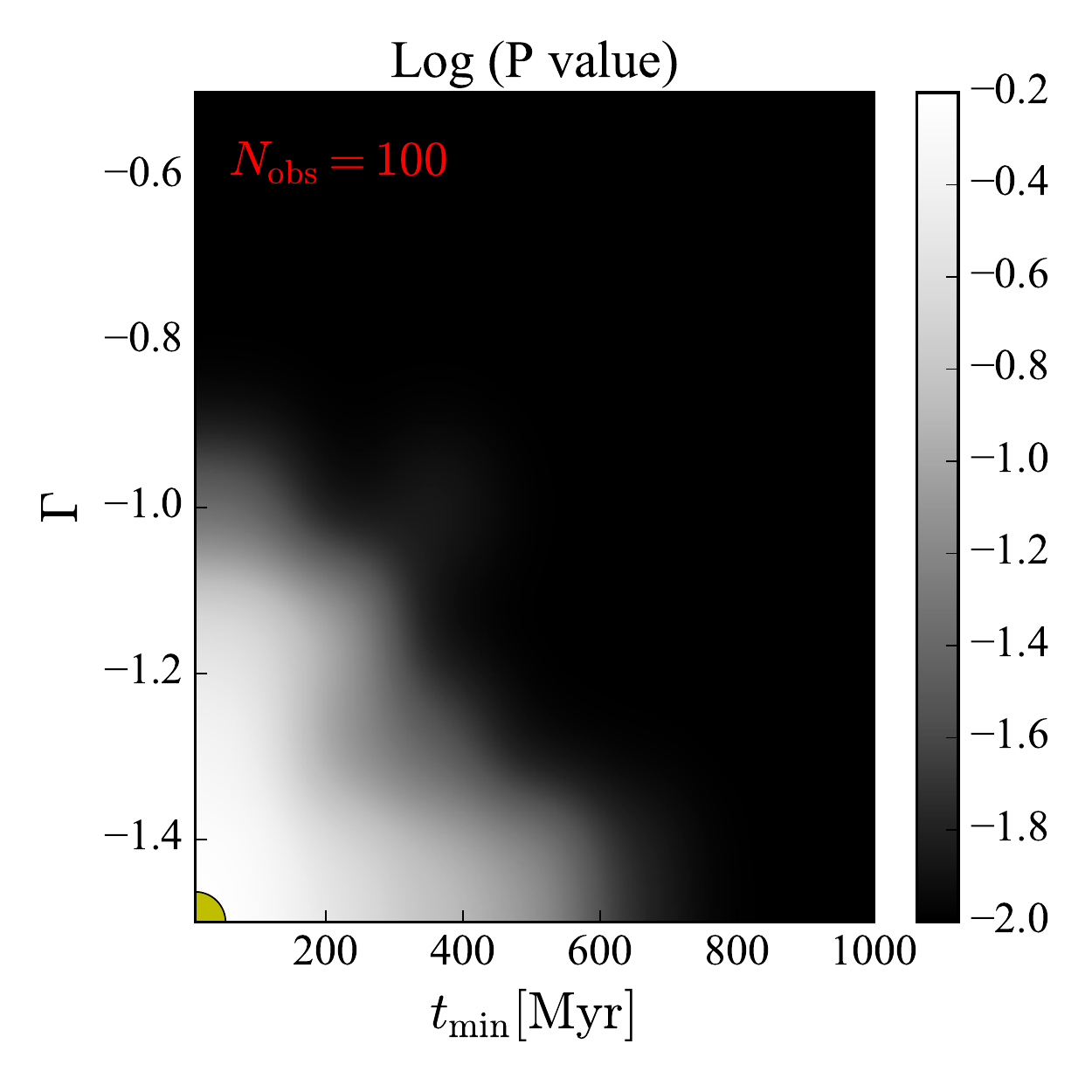}
\includegraphics[width=.33\linewidth]{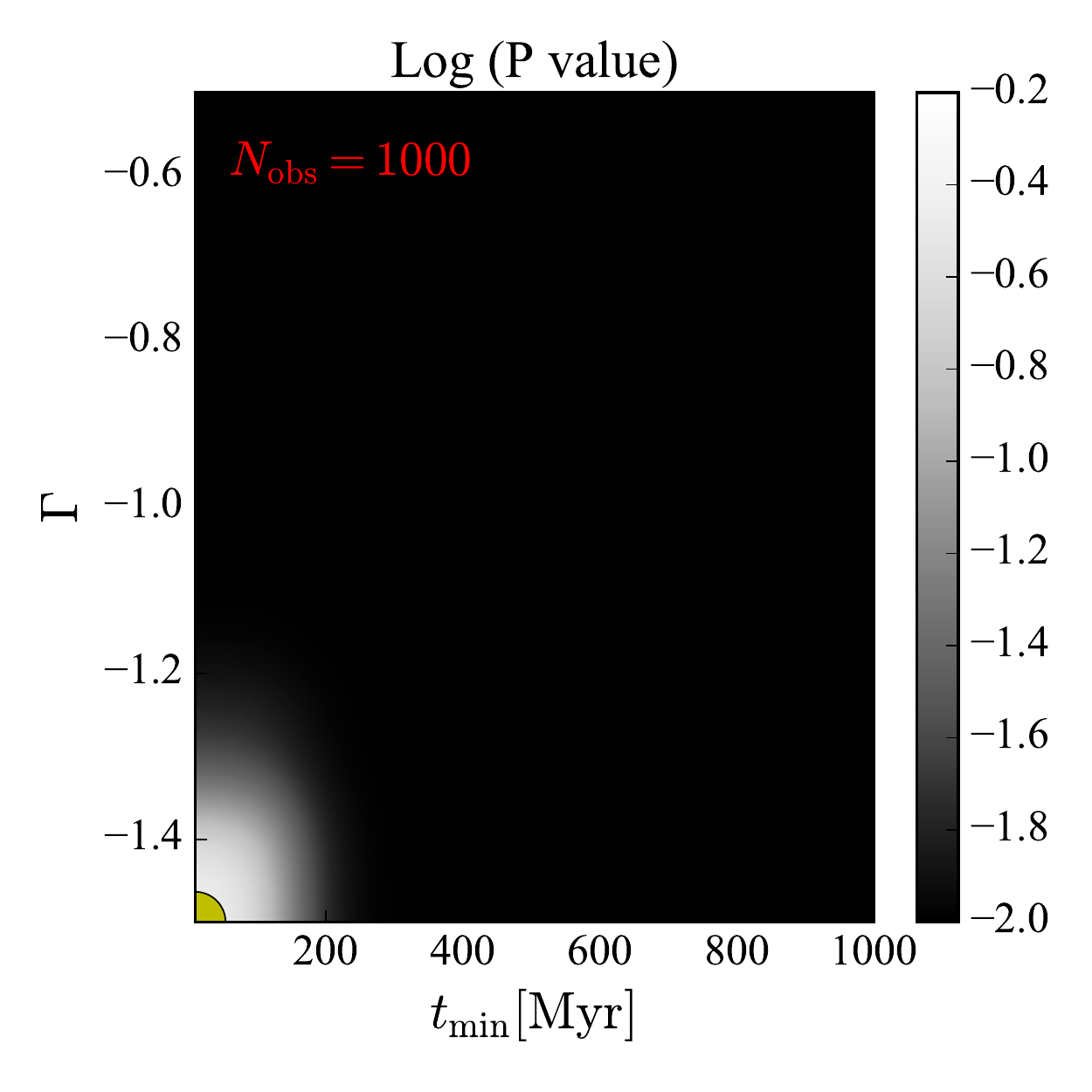}
\caption{The same as Figure \ref{f:final}, but in the space of stellar
  mass, in which we add an additional 0.3 dex uncertainty in the halo
  mass -- stellar mass relation of galaxies. The required host galaxy
  sample size is largely unchanged.}
\label{f:final_w_err}
\end{figure*}

\begin{figure}
\setlength{\tabcolsep}{1mm}
\hspace{0cm}
\includegraphics[width=\columnwidth]{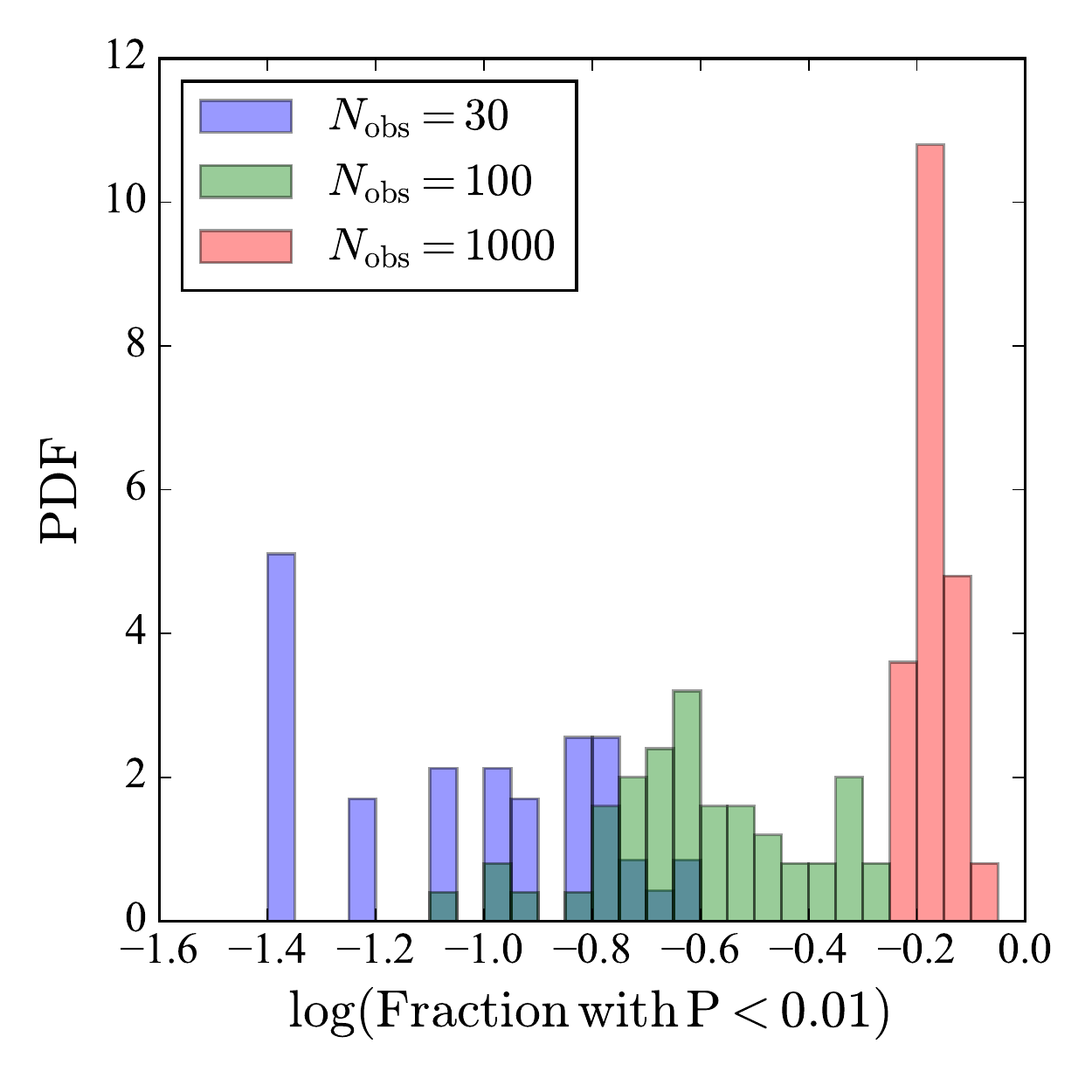}
\caption{The fraction of the $\Gamma-t_{\rm min}$ parameters space
  that can be excluded as a function of the number of observed BNS
  merger host galaxies ($N_{\rm obs}$).  This is evaluated for 50
  random [$\Gamma,t_{\rm min}$] pairs following the examples shown in
  Figure~\ref{f:final} and using a threshold of $P<0.01$.  The median
  values of the excluded fraction of the 2D parameters space are 10\%,
  25\%, and 65\% for the sample sizes of 30, 100, and 1000 BNS merger
  host galaxies, respectively.}
\label{f:overall} 
\end{figure}

\begin{figure*}
\setlength{\tabcolsep}{1mm}
\hspace{0cm}
\includegraphics[width=.33\linewidth]{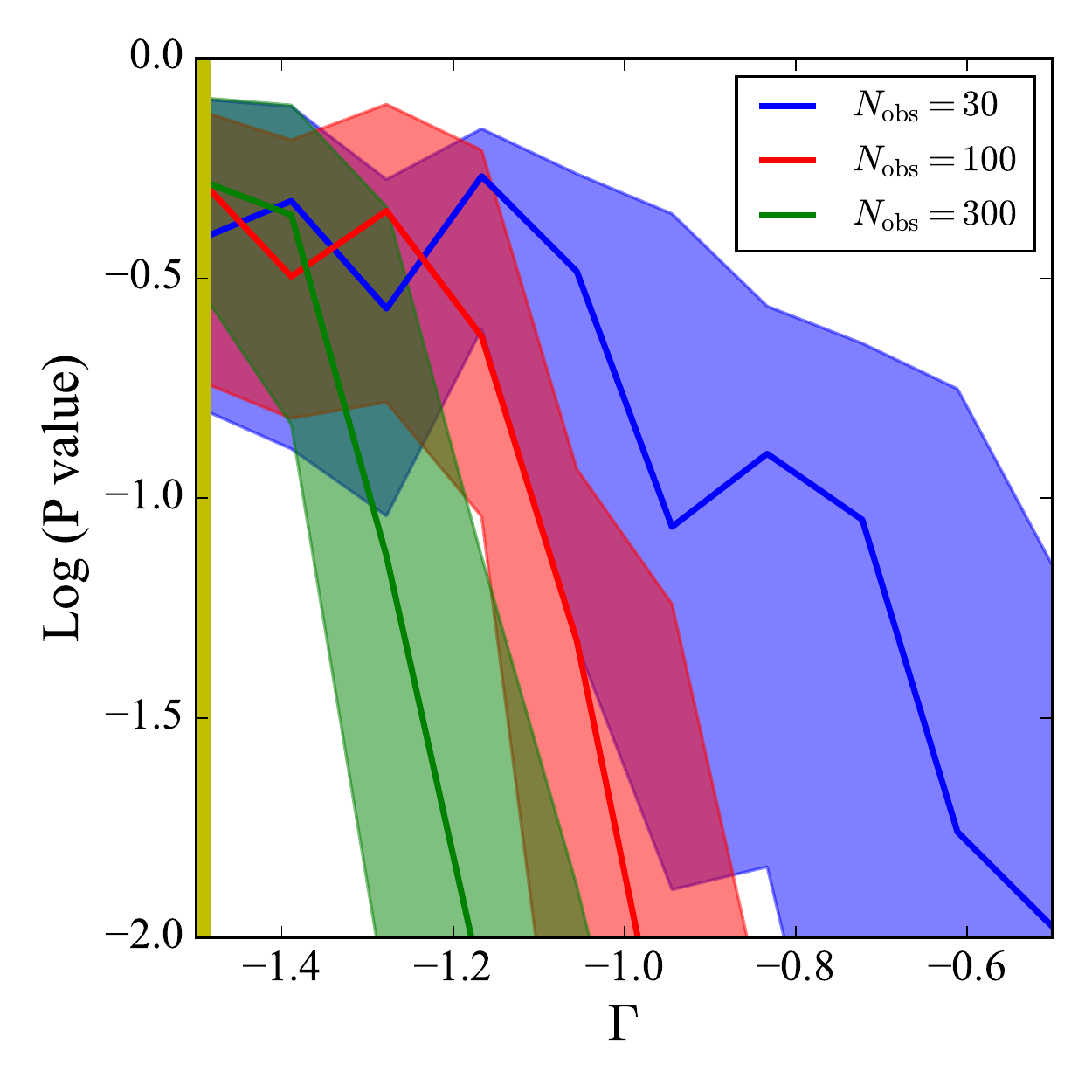}
\includegraphics[width=.33\linewidth]{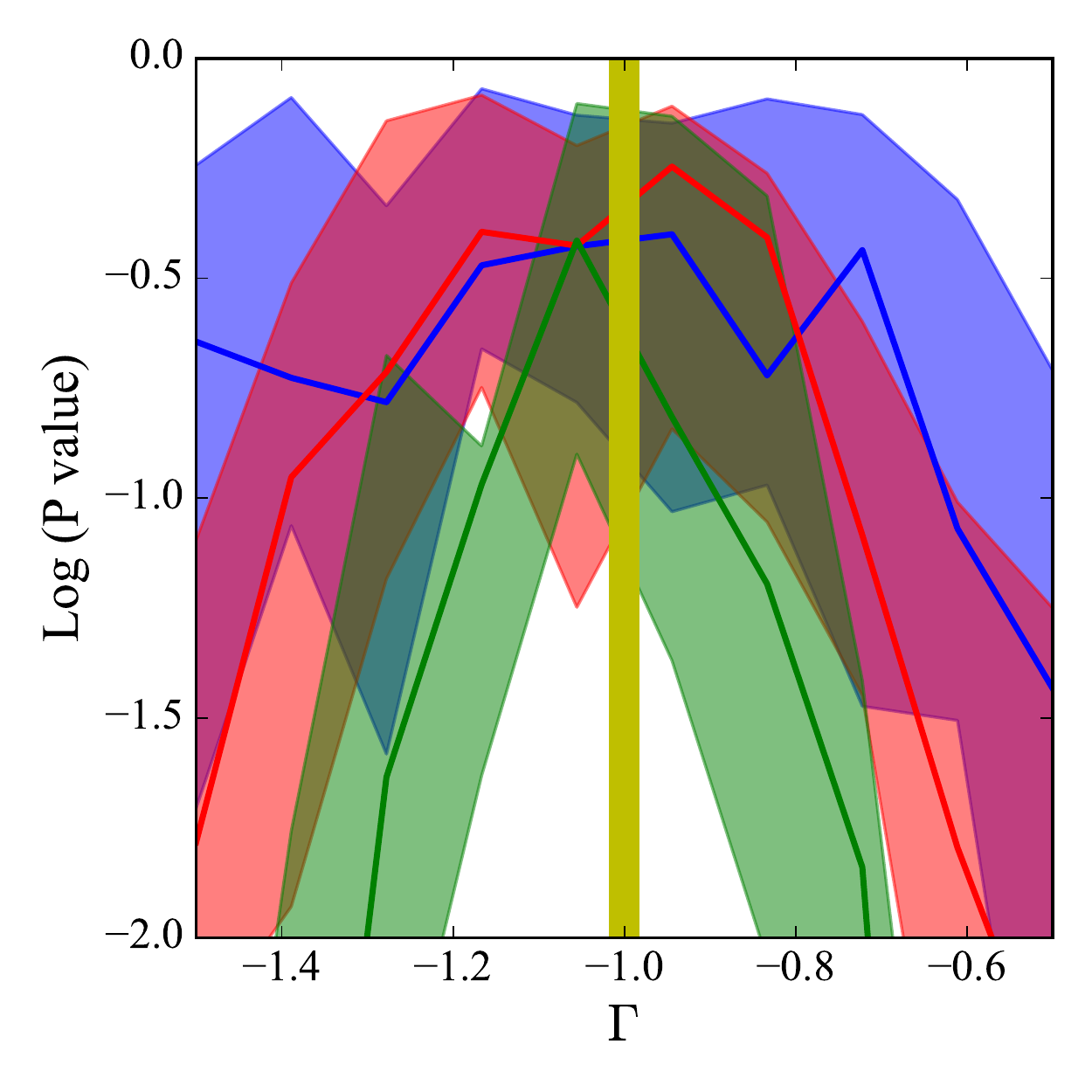}
\includegraphics[width=.33\linewidth]{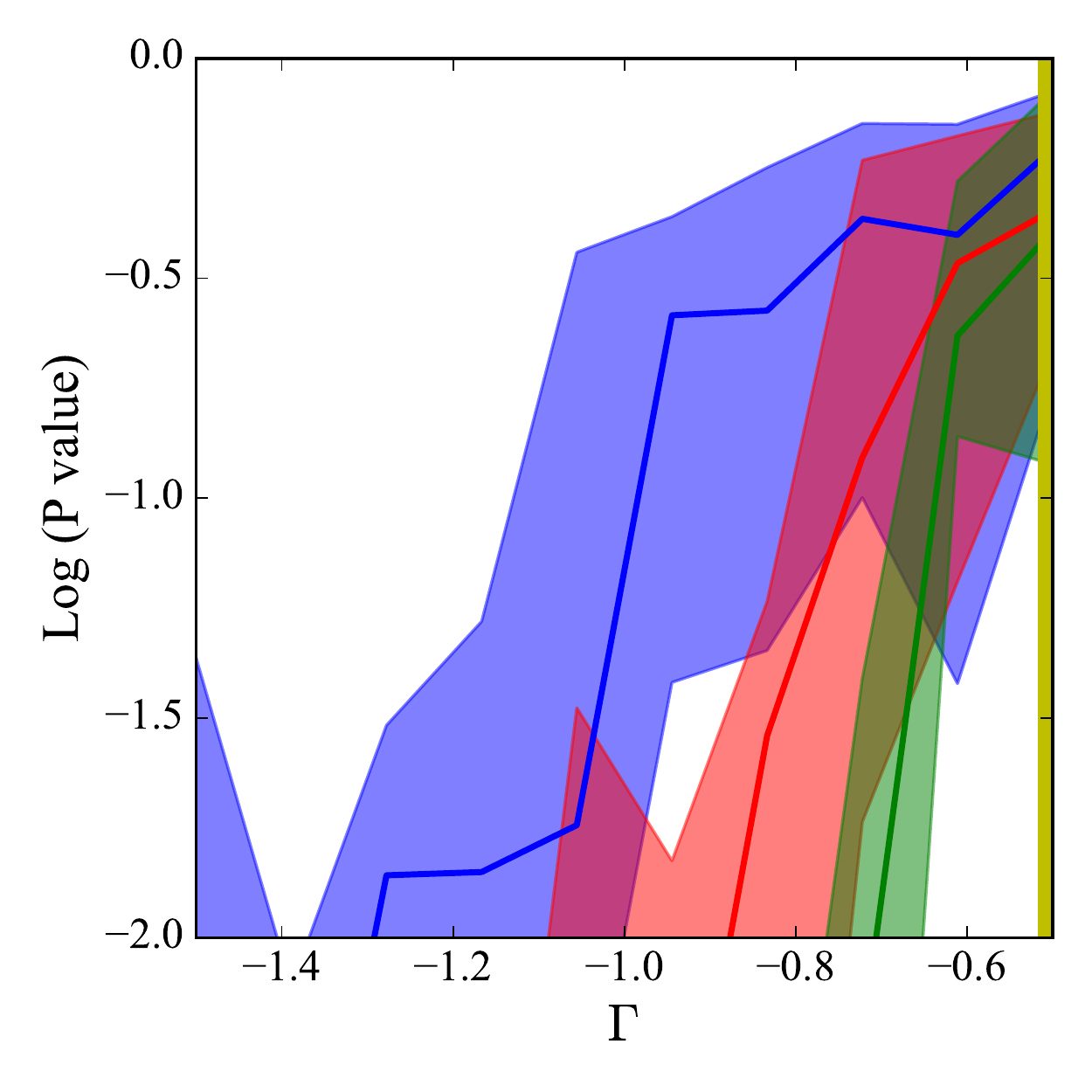}
\caption{Histograms of $P$ value in the simplified case in which we
  fix the value of $t_{\rm min}$ (to 10 Myr) and use only $\Gamma$ as
  a free parameter of the DTD.  The injected values are shown as
  yellow vertical lines and we explore host galaxy sample size of 30,
  100, and 300.  We find that a constraint on $\Gamma$ to about $30\%$
  can be achieved with a sample of about 300 host galaxies.}
\label{f:wo_tmin}
\end{figure*}

In Figure~\ref{f:final} we plot the value of $\log P$ for three
different injected DTDs and three different sample sizes (30, 100, and
1000 BNS merger host galaxies) in the plane of $\Gamma-t_{\rm min}$.
This figure uses the halo mass CDFs shown in
Figure~\ref{f:ndot_M_halo}.  We find that a sample size of
$\mathcal{O}(10^2)$ is required to begin to rule out portions of the
DTD parameter space.  With a sample size of $\mathcal{O}(10^3)$ a
significant portion of the parameter space can be ruled out, but with
a possible remaining degeneracy between $\Gamma$ and $t_{\rm min}$
depending on the true values of the parameters.  

So far we have cast our results in terms of $M_h$, but observationally
we determine the stellar mass ($M_*$) based on modeling a galaxy's
spectral energy distribution.  While there is uncertainty involved in
estimating $M_*$, the connection to $M_h$, based on abundance matching
techniques \citep{Behroozi:2013fga} is likely to dominate.  For
example, \citet{Blanchard:2017bx} determined the stellar mass of the
host galaxy of GW170817 (NGC\,4993) to be $\log
(M_*/M_{\odot})=10.90^{+0.03}_{-0.03}$ by reconstructing its star
formation history using UV to IR data.  However, the corresponding
halo mass is estimated to be $\log
(M_h/M_{\odot})=13.0^{+0.3}_{-0.3}$, with the much larger uncertainty
due to the dispersion in the $M_h-M_*$ relation
\citep{Behroozi:2013fga}.  To accommodate this uncertainty into our
analysis, we disperse the halo masses by 0.3 dex when sampling from a
given CDF, and then follow the same procedure as above to determine
$P$ values. The results of including this systematic uncertainty of
0.3 dex are equivalent to the uncertainty on the SFH of an observed
galaxy with a given stellar mass.  The inclusion of this additional
uncertainty reduces the constraining power of the observed host galaxy
sample on the DTD. However, in Figure~\ref{f:final_w_err} we show that
a sample size of $\mathcal{O}(10^3)$ can still provide significant
constraints on the DTD even when this systematic uncertainty is
included.


The overall statistics of the fraction of the $\Gamma-t_{\rm min}$
parameter space that can be ruled out with different sample sizes is
shown in Figure~\ref{f:overall}.  Here we inject 50 random
[$\Gamma,t_{\rm min}$] pairs, perform the same KS test analysis and
determine the fraction of parameter space with $P<0.01$ (i.e., the
excluded fraction).  We find that the median excluded fraction of the
parameter space is about 10\%, 25\%, and 65\% for sample sizes of 30,
100, and 1000 BNS merger host galaxies, respectively.

We note that our analysis uses a DTD with two free parameters, while
generally only $\Gamma$ is used as a free parameter, and $t_{\rm min}$
is fixed at a small value ($\sim 10$ Myr).  This simplifying
assumption is based on the notion that some binaries can merge as soon
as the second neutron star is formed.  This approach has also been
used in analyses of the DTD of Type Ia supernovae, in which a value of
$t_{\rm min}\sim 40$ Myr is often assumed (e.g.,
\citealt{Maoz:2012dg,Maoz:2014gd}), leaving only $\Gamma$ as a free
parameter.  In Figure~\ref{f:wo_tmin} we show the resulting
constraints on $\Gamma$ if we fix $t_{\rm min}=10$ Myr and repeat our
analysis.  We find that in this simplified model the value of $\Gamma$
can be determined with $\approx 30\%$ uncertainty with a reduced sample size of
$\sim 300$ BNS merger host galaxies.  Such a sample can be accumulated
in about one-third of the time compared to the requirement when both
$\Gamma$ and $t_{\rm min}$ are free parameters.

\section{Summary}

We showed how the DTD of BNS systems can be constrained through the
demographics of the host galaxies of BNS mergers detected in
gravitational waves and pinpointed through electromagnetic
observations. We focused on the case of a DTD parameterized as a power
law, although other possible DTD shapes have been proposed in the
literature (e.g., \citealt{Simonetti:2019uq}).  Our analysis is
similar to that of \citet{Zheng:2007hl}, proposed in the context of
SGRB host galaxy demographics, but with the difference that we model
the DTD with two parameters, while those authors used just $\Gamma$
and fixed $t_{\rm min}$.

Our results show that $\mathcal{O}(10^3)$ host galaxies are needed to
constrain a two-parameter DTD, although some degeneracy between
$\Gamma$ and $t_{\rm min}$ is intrinsically difficult to resolve even
with a large sample of events.  On average, about two-thirds of the
$\Gamma-t_{\rm min}$ parameter space can be ruled out with such a
sample. In the case when only $\Gamma$ is a free parameter, a sample
size of about 300 BNS merger host galaxies is sufficient for a 30\%
uncertainty on $\Gamma$.  The current range of BNS merger rates from
Advanced LIGO/Virgo Observing Runs 1 and 2 is $110-3840$ Gpc$^{-3}$
yr$^{-1}$ \citep{rate2018}, indicating that a sample of
$\mathcal{O}(10^3)$ events might be achieved within a couple of
decades at a design sensitivity of 200 Mpc, or potentially even faster
in the case of A+ \citep{A+ref}, with an expected factor of 2
increase in the BNS merger detection range.  Thus, a constraint on the
DTD using the demographics of BNS merger host galaxies can be achieved
before the advent of the third-generation GW detectors. We note that a similar approach using SGRB host galaxies is likely to take longer given an identification rate of only a few events per year \citep{Berger14}.

In an upcoming paper we will investigate a related approach to
constraining the DTD, using the measured individual SFH of BNS merger
host galaxies (rather than the mean scaling relations used here).
This is similar to the approach used by \citet{Blanchard:2017bx} for
the host galaxy of GW170817, and by \citet{Maoz:2017ex} in the context
of Type Ia SNe.  

\acknowledgements We are thankful to Enrico Ramirez-Ruiz, Or Graur,
and Evan Scannapieco for helpful discussions. This work was supported
by the National Science Foundation under grant AST14-07835 and by NASA
under theory grant NNX15AK82G.  The Berger Time-Domain Group at
Harvard is supported in part by NSF under grant AST-1714498 and by
NASA under grant NNX15AE50G. MTS is thankful to Harvard-Smithsonian
Center for Astrophysics for hospitality which made this work possible.

\end{document}